\PassOptionsToPackage{table}{xcolor}
\documentclass[11pt]{article}

\usepackage[final]{acl}

\usepackage{times}
\usepackage{latexsym}
\usepackage[utf8]{inputenc}

\usepackage[export]{adjustbox}
\usepackage[ruled]{algorithm2e}
\usepackage{float}
\usepackage[inline, shortlabels]{enumitem}
\usepackage[T1]{fontenc}
\usepackage{microtype}
\usepackage{pifont}
\usepackage[normalem]{ulem}
\usepackage{xcolor}
\usepackage{xurl}
\usepackage{graphicx}
\usepackage[section]{placeins}
\usepackage{array}
\usepackage{booktabs}
\usepackage{colortbl}
\usepackage{tabularx}
\usepackage{tabularray}
\usepackage{listings}
\usepackage[most]{tcolorbox}
\tcbuselibrary{breakable,listings,skins}
\usepackage{amsmath, amsfonts}
\usepackage{nicefrac}

\UseTblrLibrary{booktabs}

\definecolor{tblHeader}{HTML}{F4F6F8}
\definecolor{tblStripe}{HTML}{F8FAFC}
\definecolor{tblClaude}{HTML}{EEF5FF}
\definecolor{tblClaudeAlt}{HTML}{E4F0FF}
\definecolor{tblGemini}{HTML}{EFF8F5}
\definecolor{tblGeminiAlt}{HTML}{E3F1ED}
\definecolor{tblQwen}{HTML}{FFF7E8}
\definecolor{tblQwenAlt}{HTML}{FBEFD7}
\definecolor{tblBase}{HTML}{F6F7F9}
\definecolor{tblBaseAlt}{HTML}{ECEFF3}
\definecolor{tblMachine}{HTML}{EAF3FF}
\definecolor{tblOracle}{HTML}{F4ECFA}
\definecolor{tblMemory}{HTML}{EAF7ED}
\definecolor{tblPerception}{HTML}{EEF5FF}
\definecolor{tblGrounding}{HTML}{FFF3E6}
\definecolor{tblReasoning}{HTML}{EEF8F0}
\definecolor{tblSystem}{HTML}{F4EFFA}

\definecolor{NatureInk}{HTML}{243633}
\definecolor{NatureFrame}{HTML}{89A69D}
\definecolor{NatureHeader}{HTML}{D9E8E3}
\definecolor{NatureRowA}{HTML}{FBFDFB}
\definecolor{NatureRowB}{HTML}{F4F8F6}
\definecolor{PerceptionBg}{HTML}{EEF6F4}
\definecolor{GroundingBg}{HTML}{F3F0E6}
\definecolor{ReasoningBg}{HTML}{EEF3E8}
\definecolor{SystemBg}{HTML}{F4ECEC}
\definecolor{InfeasibleBg}{HTML}{ECECF4}
\definecolor{CodeBg}{HTML}{DCEAE5}
\definecolor{RuleSoft}{HTML}{B7CBC4}

\newcommand{\codepill}[1]{%
  \begingroup
  \setlength{\fboxsep}{1.3pt}%
  \colorbox{CodeBg}{\textcolor{NatureInk}{\textbf{#1}}}%
  \endgroup
}

\definecolor{promptboxbg}{HTML}{F7FCF7}
\definecolor{promptboxrule}{HTML}{3F7F45}
\definecolor{promptboxtitle}{HTML}{12351F}

\lstdefinestyle{promptlisting}{
  language={},
  basicstyle=\scriptsize\ttfamily,
  backgroundcolor=\color{promptboxbg},
  frame=none,
  breaklines=true,
  breakatwhitespace=false,
  columns=fullflexible,
  keepspaces=true,
  showstringspaces=false,
  showspaces=false,
  showtabs=false,
  numbers=none,
  aboveskip=0.6em,
  belowskip=0.8em,
}

\newtcblisting{promptbox}[1]{
  enhanced,
  breakable,
  listing only,
  listing engine=listings,
  listing options={style=promptlisting},
  colback=promptboxbg,
  colframe=promptboxrule,
  colbacktitle=promptboxrule,
  coltitle=white,
  title={\textbf{#1}},
  title after break={\textbf{#1} (continued)},
  fonttitle=\bfseries,
  boxrule=0.45pt,
  arc=1mm,
  left=0.7em,
  right=0.7em,
  top=0.45em,
  bottom=0.45em,
  before skip=0.8em,
  after skip=0.9em,
}

\makeatletter
\newcommand{\ssymbol}[1]{\@fnsymbol{#1}}
\newcommand{\romanNumeral}[1]{\expandafter\@slowromancap\romannumeral #1@}
\makeatother

\newcommand{\ours}{CUADebugger}
\newcommand{\bench}{CUAErrorBench}

\title{CUADebug: Diagnosing and Repairing Computer-Use Agent Failures}

\author{
Weijia Zhang$^{1,\S}$, Kunlun Zhu$^{1,\S}$, Zeyi Liu$^{1}$, Yinting Chen$^{1}$, Tianyi Ma$^{1}$, Jiateng Liu$^{1}$, \\
Jiaxun Zhang$^{1}$, Bingxuan Li$^{1}$, Pan Lu$^{2}$, Xiangru Tang$^{3}$, Heng Ji$^{1,\dagger}$, Jiaxuan You$^{1,\dagger}$ \\
$^{1}$University of Illinois Urbana-Champaign, Urbana, IL, USA \\
$^{2}$Stanford University, Stanford, CA, USA \\
$^{3}$Yale University, New Haven, CT, USA \\
$^{\S}$Leading authors \\
$^{\dagger}$Co-senior authors \\
\texttt{\{weijia4,kunlunz2,hengji,jiaxuan\}@illinois.edu}
}

\begin{document}
\maketitle

\begin{abstract}
Computer-use agents (CUAs) interact with graphical interfaces through screenshots and low-level mouse and keyboard actions, yet the causal error may precede the terminal failure. We present \textbf{CUADebug}, a framework for localizing root causes in CUA trajectories and guiding re-execution. \textbf{CUADebug} includes a five-category, 30-subtype taxonomy; \bench{}, a benchmark of 204 failed OSWorld trajectories with human root-cause annotations; and \ours{}, a ReAct-style agent for root-cause analysis (RCA). \ours{} iteratively selects trajectory steps, inspects paired before/after screenshots and action traces, and submits a structured diagnosis containing the causal step, taxonomy label, grounded evidence, and correction. \ours{} performs RCA without per-trajectory human intervention; human annotations are used to evaluate RCA predictions and, in controlled re-rollout comparisons, to fix restart points. Task reasoning and control is the largest annotated failure category (110/204). \ours{} improves L2 and Tag+Step Exact across three debugger backbones on the Claude-agent split; with Gemini 2.5 Pro, Tag+Step Exact rises from 11.1\% to 19.4\%. Single re-execution improves failure recovery from 13.89\% to 29.86\% (overall 61.77\% to 68.14\%); controlled continual re-execution improves it from 12.50\% to 25.69\% (overall 61.22\% to 66.48\%).
Project page: \linebreak \href{https://cuadebug.github.io/}{\texttt{cuadebug.github.io}}.
\end{abstract}

\section{Introduction}

Computer-use agents (CUAs) execute tasks in desktop, web, and mobile interfaces through screenshots and mouse or keyboard actions. Compared with text-only or tool-only agents, CUA agents must jointly solve visual perception, spatial grounding, low-level interaction mechanics, and long-horizon state tracking. When a CUA fails, the terminal file, chart, or UI state may be a downstream symptom of an earlier perception, grounding, reasoning, or environment error. Benchmarks such as OSWorld~\citep{osworld}, WebArena~\citep{webarena}, VisualWebArena~\citep{visualwebarena}, Mind2Web~\citep{mind2web}, AndroidWorld~\citep{androidworld}, and Windows Agent Arena~\citep{windowsarena} measure end-to-end task success, but they do not localize the causal step in a failed trajectory.

\begin{figure*}[t]
\centering
\includegraphics[width=\textwidth,trim=12bp 12bp 9bp 15bp,clip]{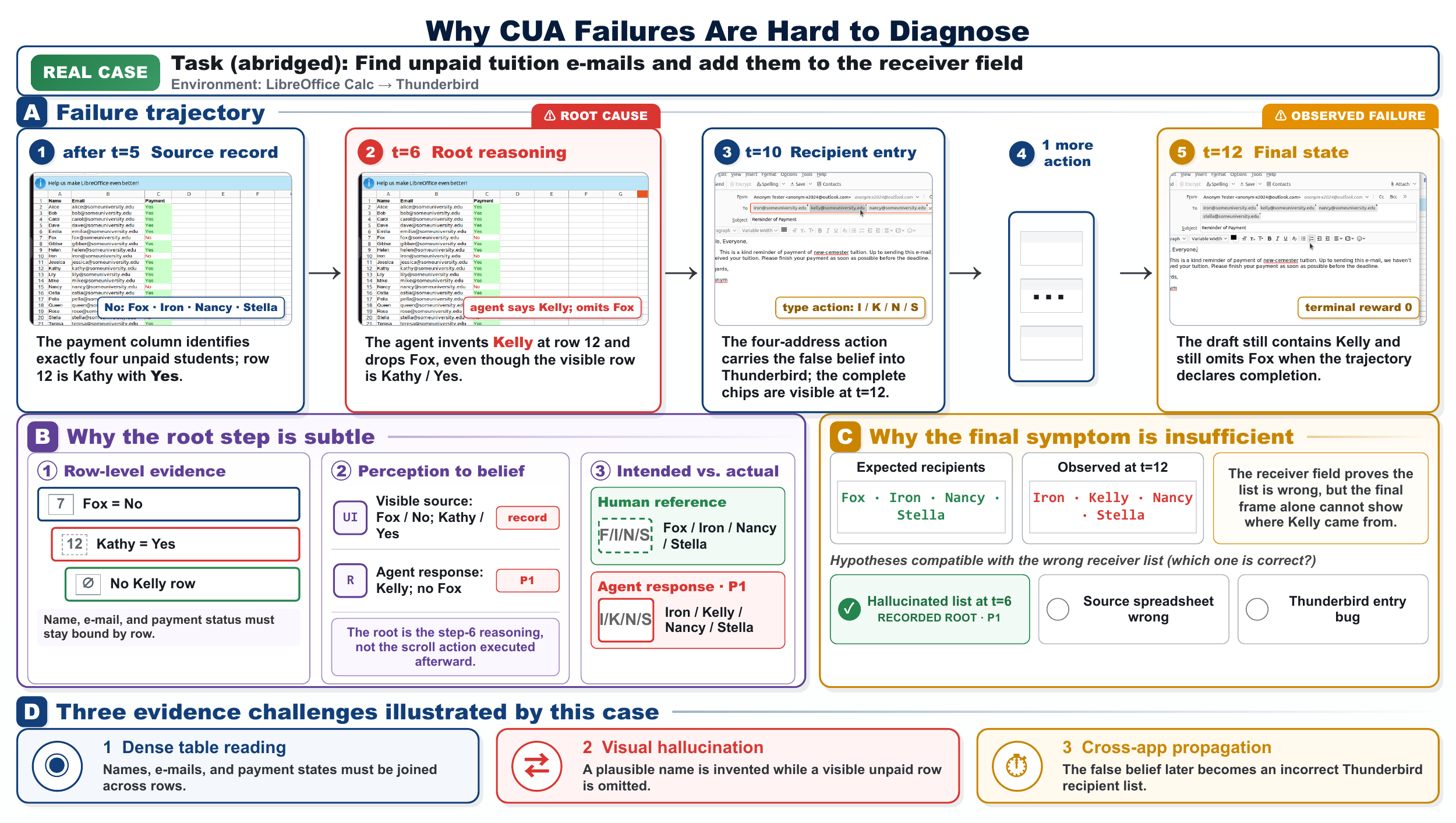}
\caption{Why CUA failures require root-cause diagnosis rather than terminal-state inspection.}
\label{fig:teaser}
\end{figure*}

Existing agent evaluations report end-to-end task success without localizing the step or cause of failure. Error taxonomies and failure-attribution studies identify recurring agent mistakes~\citep{agenterrorbench,mast,agentic_faults,whowhen,raffles,agentfail}, but CUA root-cause analysis must localize a causal step from evidence distributed across screenshots, spatial targets, low-level actions, and temporal screen transitions. Figure~\ref{fig:teaser} illustrates this temporal gap: the earliest causal error can precede the terminal failure. To address this challenge, we develop \ours{}, an automatic ReAct-style debugger that uses an iterative tool-use loop to inspect selected trajectory steps using paired before/after screenshots and action traces, then produces a structured root-cause analysis (RCA) record for re-execution.

We evaluate CUADebug through three research questions. First, where do CUA agents fail? We characterize these failures with a CUA-specific taxonomy and human annotations of failed OSWorld trajectories. Second, does \ours{}'s tool-augmented trajectory inspection improve RCA accuracy over a prompt-only baseline? We compare \ours{} with a prompt-only baseline that returns the same structured RCA fields, using human annotations as ground truth for taxonomy agreement and root-step localization. Third, can RCA guide re-execution? We evaluate whether diagnostic context improves task completion under single and continual re-rollout protocols.

To evaluate the alignment of an LLM-based RCA judge with human experts, we introduce \bench{}, a human-annotated benchmark of failed OSWorld trajectories. For each failed trajectory, human experts annotate the root-cause step, L1/L2 taxonomy label, supporting evidence, corrective strategy, and confidence. \bench{} enables direct comparison between LLM-generated RCA and human expert judgments at both the taxonomy and root-step levels. Our benchmark contains 204 usable annotations across Claude 4.5 Sonnet, Gemini 2.5 Pro, and Qwen 3.5 trajectory sources.

\begin{figure*}[!t]
\centering
\includegraphics[width=\textwidth,trim=16bp 11bp 42bp 12bp,clip]{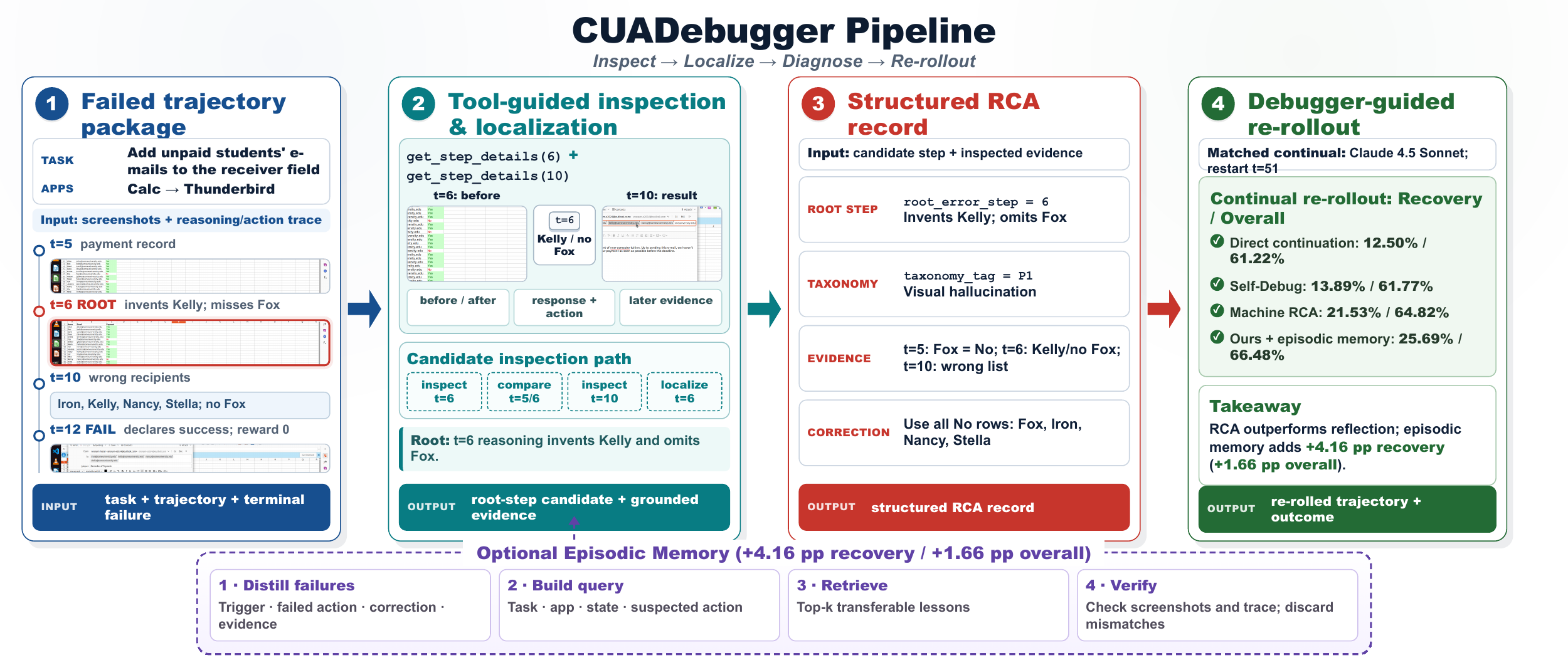}
\caption{\ours{} inspects failed trajectories, produces structured RCA, retrieves relevant memories, and guides re-rollout.}
\label{fig:pipeline}
\end{figure*}

\ours{} implements this diagnosis process with two CUA-specific tools (Figure~\ref{fig:pipeline}). Within a ReAct~\citep{react} loop, the step-inspection tool returns paired before/after screenshots, actions, reasoning, and execution status for selected steps, while the submission tool records a coarse-to-fine taxonomy label, grounded evidence, a concrete correction, confidence, and per-step summaries for re-rollout. When episodic memory is enabled, \ours{} retrieves diagnoses distilled from prior debugging episodes as candidate evidence for the current RCA. CUADebugger passes the predicted root-cause step, grounded evidence, and correction to the re-rollout agent. Re-rollout task completion then measures whether this diagnostic context supports repair.

\begin{figure}[!t]
\centering
\includegraphics[width=\columnwidth,trim=13bp 4bp 3bp 5bp,clip]{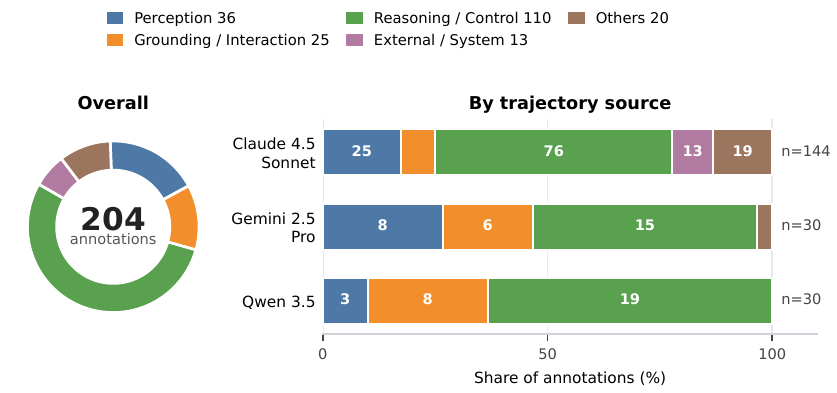}
\caption{\bench{} annotation distribution by trajectory source.}
\label{fig:bench-stats}
\end{figure}

First, task reasoning and control accounts for the largest share of human-annotated failures (110/204), while the remaining cases span perception, grounding/interaction, external/system, and Others (Figure~\ref{fig:bench-stats}). Second, \ours{} improves L2 and Tag+Step Exact over the prompt-only baseline for all three debugger backbones on the Claude-agent split; with Gemini 2.5 Pro, Tag+Step Exact increases from 11.1\% to 19.4\%. Third, in the single re-execution package comparison, \ours{} achieves 29.86\% task completion versus 13.89\% for history-only continuation (68.14\% versus 61.77\% Overall Acc); under the more controlled continual protocol, it achieves 25.69\% versus 12.50\% (66.48\% versus 61.22\% Overall Acc), while human-oracle guidance reaches 29.17\% (67.87\% Overall Acc).

We make three contributions. First, we characterize CUA failures with a root-cause taxonomy comprising five categories and 30 subtypes (Table~\ref{tab:taxonomy-l1}). Second, we introduce \bench{}, a benchmark with human root-cause annotations for 204 failed OSWorld trajectories. Third, we develop \ours{}, an automatic ReAct-style RCA agent that actively selects trajectory steps for multimodal inspection, submits structured diagnoses, and optionally retrieves prior debugging episodes; we evaluate its RCA against human annotations and its corrections through re-execution.

\section{CUAErrorBench}

We construct \bench{} from failed OSWorld trajectories generated by three CUA configurations. Each trajectory contains a task instruction, visual observations, screenshots, model reasoning traces, low-level actions, rewards, and terminal status. \bench{} contains 204 annotated trajectories: 144 generated by Claude 4.5 Sonnet with a 50-step budget, 30 by Gemini 2.5 Pro, and 30 by Qwen 3.5. Figure~\ref{fig:bench-stats} summarizes the category distribution across trajectory sources; task reasoning and control is the largest category (110/204).

The annotation pipeline stores debugger-generated RCA proposals alongside human annotations, which serve as ground truth for evaluating taxonomy agreement and root-step localization. For each failed trajectory, annotators record five fields: the root-cause step, L1/L2 taxonomy label, supporting evidence, corrective strategy, and confidence. Across all 204 cases ($N=204$), two annotators separately recorded first-pass annotations before discussion. The pre-adjudication exact-agreement rates were 85.8\% for the root-cause step, 94.6\% for the L1 category, and 85.8\% for the L2 subtype. Disagreements were adjudicated to produce one ground-truth annotation per trajectory. Appendix~\ref{app:human-annotation-details} provides the annotation instructions and consent procedure; Appendix~\ref{app:annotation-case} presents a complete annotated case.

\paragraph{Benchmark curation and annotation protocol.}
\label{sec:bench-protocol}
To verify each RCA proposal, annotators inspect the task instruction, terminal failure signal, action and reasoning trace, and before/after screenshots. They verify the terminal failure and trace backward to the earliest step that introduces a new causal error, excluding steps that merely inherit an earlier mistake. Supporting evidence must cite concrete trajectory observations, such as the visible UI state, selected element, executed action, or a mismatch between the agent's stated intention and the resulting screen state. The correction specifies an alternative action at the root-cause step, and confidence reflects whether competing explanations remain plausible.

\paragraph{Taxonomy labels.}
The two-level CUA taxonomy provides a shared label space for automatic RCA, human evaluation, and memory indexing. The top-level label identifies the causal module, while the subtype specifies the failure pattern used for fine-grained evaluation and memory indexing. Table~\ref{tab:taxonomy-l1} defines the five top-level categories, and Appendix~\ref{app:taxonomy} lists all 30 subtypes.

\begin{table}[!h]
\centering
\tiny
\setlength{\tabcolsep}{2.5pt}
\renewcommand{\arraystretch}{1.05}
\begin{tcolorbox}[
    colback=white,
    colframe=NatureFrame,
    boxrule=0.55pt,
    arc=2pt,
    left=3pt,
    right=3pt,
    top=3pt,
    bottom=3pt
]
\arrayrulecolor{RuleSoft}
\rowcolors{2}{NatureRowA}{NatureRowB}
\begin{tabularx}{\linewidth}{@{}
    >{\centering\arraybackslash}p{0.065\linewidth}
    >{\raggedright\arraybackslash}p{0.150\linewidth}
    >{\raggedright\arraybackslash}p{0.270\linewidth}
    >{\raggedright\arraybackslash}X
@{}}
\rowcolor{NatureHeader}
\textbf{Code} & \textbf{Error module} & \textbf{Diagnostic question} & \textbf{Typical evidence} \\
\midrule

\rowcolor{PerceptionBg}
\codepill{P} & \textbf{Perception} &
Did the agent misunderstand what was visible in the observation? &
Hallucinated or missed UI content, OCR mistakes, cross-region binding errors, semantic misreadings. \\[1pt]

\rowcolor{GroundingBg}
\codepill{G} & \textbf{Grounding and Interaction} &
Did the agent know the intended operation but execute it on the wrong target or with the wrong mechanics? &
Wrong coordinates or element, hidden or disabled target, incorrect click, drag, gesture, or text-entry behavior. \\[1pt]

\rowcolor{ReasoningBg}
\codepill{R} & \textbf{Task Reasoning and Control} &
Did the agent choose, maintain, or revise the wrong plan? &
Constraint violations, missing subgoals, action-intent mismatch, memory loss, progress misjudgment, failed self-correction. \\[1pt]

\rowcolor{SystemBg}
\codepill{S} & \textbf{External/System} &
Did the environment, tool, or benchmark setup prevent otherwise valid progress? &
Rendering or timing failures, unexpected system dialogs, resource limits, tool/API failures, benchmark artifacts. \\[1pt]

\rowcolor[HTML]{F1F1FA}
\codepill{O} & \textbf{Others} &
Does the failure fall outside the four execution modules above? &
Cases not attributable to P/G/R/S. \\

\end{tabularx}
\rowcolors{2}{}{}
\arrayrulecolor{black}
\end{tcolorbox}
\caption{Top-level CUA error taxonomy used in \textit{CUAErrorBench}.}
\label{tab:taxonomy-l1}
\end{table}

\section{Method}

\ours{} diagnoses and repairs CUA failures through three components: tool-augmented RCA, optional episodic-memory retrieval, and debugger-guided re-rollout.

\subsection{Root-Cause Analysis}
\label{sec:method-rca}

RCA localizes the earliest causal step in a failed trajectory and produces a structured diagnosis. \ours{} runs a ReAct~\citep{react} loop that iteratively selects trajectory steps and invokes CUA-specific tools to inspect their multimodal evidence. The RCA agent receives the task instruction, trajectory metadata, terminal failure step $F$, L1/L2 taxonomy definitions, and output schema; retrieved memories are added as candidate evidence only when memory is enabled. At each turn, the agent selects the next step to inspect, retains a textual observation note, and removes older screenshots from the context window.

\paragraph{Multimodal step-inspection tool.}
\texttt{get\_step\_details}$(n)$ returns the action, reasoning trace, execution error, reward, and screenshots immediately before and after step $n$. Comparing the before/after screenshots with the agent's stated intention reveals perception, grounding, and outcome-misinterpretation errors.

\paragraph{Structured root-cause submission tool.}
The agent terminates by calling \texttt{finish}$(\cdot)$. The \texttt{finish} schema requires the root-cause step $N$, an L1/L2 taxonomy label, grounded evidence, a concrete correction, confidence in $[0,1]$, and a summary of each inspected step. These summaries, together with the predicted root-cause step, taxonomy label, evidence, and correction, form the diagnostic context passed to re-rollout (Section~\ref{sec:method-rerollout}).

\paragraph{Memory retrieval.}
When episodic memory is enabled, the RCA prompt includes the top-$k$ retrieved memories as candidate evidence; the agent compares each trigger condition with the current trajectory and discards mismatches. Appendix~\ref{app:rca-prompt-tools} provides the RCA prompt template, memory-retrieval block, and tool schemas.

\subsection{Error Taxonomy Module}

The error taxonomy module makes RCA a coarse-to-fine diagnosis. For each candidate root-cause step, the debugger selects one of five top-level modules (Table~\ref{tab:taxonomy-l1}) and then a subtype within that module (Appendix~\ref{app:taxonomy}). The top-level label identifies the causal source used to guide re-rollout, while the subtype supports memory indexing and fine-grained evaluation.

\subsection{Multimodal Trajectory Inspection}

\ours{} analyzes the before/after screenshots and corresponding action trace from each inspected step as paired causal evidence. For each inspected step, the debugger records the pre-action screen state, intended UI operation, executed low-level action, and resulting screen state. These paired observations reveal errors such as selecting the wrong chart subtype, clicking a nearby control, or assuming that a dialog changed state when the resulting screenshot shows otherwise. The debugger uses the resulting step summaries for RCA, memory distillation, and re-rollout.

\subsection{Episodic Memory and Retrieval-Augmented Debugging}
\label{sec:method-memory}

Episodic memory converts individual trajectories into reusable debugging knowledge. For each failed trajectory, \ours{} stores an episode containing task and application metadata, local error context, inferred agent intention, screen outcome, taxonomy label, and RCA output. During retrieval, \ours{} returns the distilled rule together with the supporting trajectory evidence stored in the episode.

Retrieval-augmented debugging exposes this memory through a tool-use interface. From each episode, \ours{} distills a memory record containing a title, trigger condition, failed and corrected actions, taxonomy label, distinguishing feature, supporting evidence, and source-episode references. Memories are indexed by application identifier, taxonomy label, trigger condition, failed action, and memory text. During RCA or re-rollout, the debugger or acting agent retrieves similar failures as candidate evidence and compares their trigger conditions and distinguishing features with the current screenshots and action trace. At cold start, the memory is empty; it then grows from successful RCA outputs and contrastive failure/success trajectory pairs. We evaluate the contribution of memory through a staged comparison between Machine RCA without retrieval and the memory-enabled variant.

Algorithm~\ref{alg:rag} summarizes memory construction and retrieval for RCA and re-rollout.

\begin{algorithm}[t]
\small
\DontPrintSemicolon
\KwIn{Prior debugging episodes $\mathcal{D}_{mem}$, target trajectory $\tau$, retrieval size $k$, mode $m \in \{\mathrm{RCA}, \mathrm{re\mbox{-}rollout}\}$}
\KwOut{An RCA report or re-rollout guidance}
$M \leftarrow \emptyset$\;
\For{$\tau_i \in \mathcal{D}_{mem}$}{
    $r_i \leftarrow \textsc{RunRCA}(\tau_i)$\;
    \If{$r_i$ contains a root step, tag, evidence, and correction}{
        $e_i \leftarrow \textsc{BuildEpisode}(\tau_i, r_i)$\;
        \If{a corrected re-rollout $\tau_i^{+}$ exists}{
            attach the failure/success contrast from $(\tau_i, \tau_i^{+})$ to $e_i$\;
        }
        $M.\textsc{Add}(\textsc{DistillMemory}(e_i))$\;
    }
}
$q \leftarrow \textsc{BuildQuery}(\tau, m)$\;
$L \leftarrow \texttt{memory\_query}(M, q, k)$\;
$C \leftarrow \{\ell \in L : \textsc{Match}(\ell, \tau)\}$\;
\If{$m=\mathrm{RCA}$}{
    \Return $\textsc{Diagnose}(\tau, C)$\;
}
\Return $\textsc{BuildRepairGuidance}(\tau, C)$\;
\caption{Building and querying episodic memory for RCA and re-rollout.}
\label{alg:rag}
\end{algorithm}
\FloatBarrier

\subsection{Debugger-Guided Re-Rollout}
\label{sec:method-rerollout}
Re-rollout measures whether diagnostic context improves subsequent task execution. Both the direct baseline and the debugger-guided method replay the original OSWorld trajectory to the same cutoff state and continue with the same remaining step budget. With state and budget fixed, the debugger-guided condition adds the previous trajectory summary, inspected-step summaries, and a repair recipe containing the predicted root-cause step, taxonomy label, evidence, and correction to the acting prompt. In the retrieval-enabled variant, the acting agent can additionally query episodic memory for similar prior failures before choosing its next action. Appendix~\ref{app:rerollout-prompts} lists the re-rollout prompt variants and visual side-channel tools.

We measure whether diagnostic context supports behavioral repair during re-rollout using task completion and root-error repair.

\section{Experiments}

\subsection{Evaluation Protocol}

We evaluate CUA debugging along three research questions. First, failure analysis asks where CUA agents fail by measuring the human-labeled distribution of root-cause modules across trajectory sources. Second, RCA evaluation asks whether \ours{} improves over a naive prompt baseline and matches human root-cause judgments. Third, re-rollout evaluation asks whether RCA can guide subsequent execution. Unless otherwise stated, human annotations are the reference labels. We compute Acc over the failed trajectories sent to re-execution and Overall Acc over all 361 tasks, counting both initial successes and recovered failures.

\subsection{Failure Analysis}

Failure analysis uses the human labels in \bench{} to characterize where CUA agents fail. We report top-level taxonomy distributions by agent trajectory source in Figure~\ref{fig:bench-stats}. This analysis answers a descriptive question rather than a model-ranking question: it provides the reference view of CUA failure modes against which model debugger behavior is later compared.

\subsection{RCA Evaluation}

The RCA evaluation reports both taxonomy and localization accuracy. We compare two RCA methods. The \emph{Naive Baseline} receives the task, full trajectory package, and taxonomy definitions, and produces the RCA record in one pass. \ours{} receives the same task metadata and taxonomy definitions but must use the step-inspection and structured-submission tools described in Section~\ref{sec:method-rca} to inspect step details before producing the same RCA record. Table~\ref{tab:rca} reports both methods. L1 accuracy measures top-level taxonomy agreement with the human annotation, while L2 accuracy measures exact subtype agreement. Step Exact measures exact root-cause-step match, and Step $\pm 2$ gives credit when the predicted root-cause step is within two steps of the human-labeled root-cause step. Tag+Step Exact requires both L2 subtype match and exact root-cause-step match.

\begin{table}[t]
\centering
\tiny
\setlength{\tabcolsep}{2.0pt}
\resizebox{0.96\columnwidth}{!}{%
\rowcolors{2}{white}{tblStripe}
\begin{tabular}{@{}lllrrrrr@{}}
\toprule
\rowcolor{tblHeader}
\shortstack{Agent\\traj.} & \shortstack{RCA\\method} & Debugger & L1 & L2 & \shortstack{Step\\Exact} & \shortstack{Step\\$\pm$2} & \shortstack{Tag+\\Step} \\
\midrule
Claude 4.5 & Naive & Gemini 2.5 Pro & 50.7 & 29.9 & 20.8 & 38.2 & 11.1 \\
Claude 4.5 & \textbf{CUADebug} & Gemini 2.5 Pro & 53.5 & 36.8 & 29.2 & 47.9 & \textbf{19.4} \\
\cmidrule(lr){2-8}
Claude 4.5 & Naive & Qwen 3.5 & 44.4 & 19.4 & 23.6 & 34.7 & 7.6 \\
Claude 4.5 & \textbf{CUADebug} & Qwen 3.5 & 57.6 & 33.3 & 25.0 & 46.5 & \textbf{14.6} \\
\cmidrule(lr){2-8}
Claude 4.5 & Naive & Claude 4.5 & 49.3 & 14.6 & 22.2 & 38.2 & 4.9 \\
Claude 4.5 & \textbf{CUADebug} & Claude 4.5 & 61.1 & 31.3 & 29.2 & 45.1 & \textbf{15.3} \\
\midrule
Gemini 2.5 Pro & \textbf{Naive} & Qwen 3.5 & 46.7 & 20.0 & 26.7 & 50.0 & \textbf{6.7} \\
Gemini 2.5 Pro & CUADebug & Qwen 3.5 & 43.3 & 20.0 & 23.3 & 40.0 & 3.3 \\
\cmidrule(lr){2-8}
Gemini 2.5 Pro & Naive & Claude 4.5 & 53.3 & 23.3 & 26.7 & 50.0 & 10.0 \\
Gemini 2.5 Pro & \textbf{CUADebug} & Claude 4.5 & 43.3 & 23.3 & 33.3 & 50.0 & \textbf{16.7} \\
\cmidrule(lr){2-8}
Gemini 2.5 Pro & Naive & Gemini 2.5 Pro & 50.0 & 20.0 & 30.0 & 40.0 & \textbf{10.0} \\
Gemini 2.5 Pro & CUADebug & Gemini 2.5 Pro & 53.3 & 16.7 & 40.0 & 50.0 & \textbf{10.0} \\
\midrule
Qwen 3.5 & Naive & Gemini 2.5 Pro & 53.3 & 23.3 & 33.3 & 43.3 & 10.0 \\
Qwen 3.5 & \textbf{CUADebug} & Gemini 2.5 Pro & 53.3 & 36.7 & 46.7 & 63.3 & \textbf{16.7} \\
\cmidrule(lr){2-8}
Qwen 3.5 & Naive & Qwen 3.5 & 46.7 & 20.0 & 16.7 & 33.3 & 3.3 \\
Qwen 3.5 & \textbf{CUADebug} & Qwen 3.5 & 43.3 & 20.0 & 20.0 & 36.7 & \textbf{13.3} \\
\cmidrule(lr){2-8}
Qwen 3.5 & \textbf{Naive} & Claude 4.5 & 56.7 & 30.0 & 30.0 & 36.7 & \textbf{10.0} \\
Qwen 3.5 & CUADebug & Claude 4.5 & 46.7 & 36.7 & 26.7 & 36.7 & 10.0 \\
\bottomrule
\end{tabular}
\rowcolors{2}{}{}
}
\caption{RCA accuracy against human annotations (percent). Claude-agent results use $N=144$; Gemini- and Qwen-agent results use $N=30$. CUADebug denotes our tool-augmented RCA method; Tag+Step requires exact subtype and exact root-cause-step match.}
\label{tab:rca}
\end{table}

\subsection{Single Re-Rollout Evaluation}

Single re-rollout isolates whether a diagnosis can support behavioral repair. The rollout agent restarts one step before the root-cause step and runs with a fixed step and token budget. The Machine RCA condition uses the machine-predicted root-cause step and machine-generated diagnosis without memory retrieval to test the automatic RCA-to-repair pipeline, while the baseline, self-debug, human-oracle, and our method conditions use the human-labeled root-cause step to fix the restart point and compare diagnostic context. This controls for localization error: if the restart step itself is wrong, the re-rollout may be uninformative regardless of the debugging method. Because the rows differ in diagnostic prompt format, Table~\ref{tab:single-rerollout} should be read as a re-execution-package comparison rather than a clean ablation of RCA source quality. The main root-error metric is New+Fixed: a post-analysis judge compares the original human annotation, including root-cause step, evidence, and correction, against the new trajectory; the first number counts cases where the original critical error is fixed but a new critical error appears, and the second counts cases where the original critical error is fixed without a new critical error. When the budget reaches a task-level outcome we additionally report Re-rollout Acc as task completion rate. The agent trajectory and rollout agent are fixed to Claude 4.5 Sonnet, and Machine RCA diagnoses are produced by Gemini 2.5 Pro.

\begin{table}[t]
\centering
\scriptsize
\setlength{\tabcolsep}{3pt}
\resizebox{\columnwidth}{!}{%
\begin{tabular}{@{}p{0.18\linewidth}p{0.22\linewidth}rrrrrr@{}}
\toprule
\multicolumn{2}{c}{Condition} & \multicolumn{3}{c}{Cost} & \multicolumn{2}{c}{Task} & \multicolumn{1}{c}{Root error} \\
\cmidrule(lr){1-2}\cmidrule(lr){3-5}\cmidrule(lr){6-7}\cmidrule(l){8-8}
Method & RCA source & \shortstack{Avg.\\turns} & \shortstack{Input\\tok.} & \shortstack{Output\\tok.} & \shortstack{Acc.\\(\%)} & \shortstack{Overall\\Acc. (\%)} & \shortstack{New+\\Fixed} \\
\midrule
Baseline & none & 28.01 & 272.7K & 3.5K & 13.89 & 61.77 & 20+11 \\
Self-debug & self & 31.53 & 324.4K & 3.9K & 15.28 & 62.33 & 22+14 \\
Machine RCA & machine RCA & 23.03 & 290.8K & 2.9K & 28.47 & 67.59 & 40+25 \\
Human oracle & human annotation & 22.37 & 296.9K & 3.7K & \textbf{31.94} & \textbf{68.98} & 46+29 \\
\textbf{Our method} & RCA + memory & 28.28 & 319.1K & 2.9K & 29.86 & 68.14 & 43+24 \\
\bottomrule
\end{tabular}
}
\caption{Single re-rollout results on Claude 4.5 Sonnet trajectories. Overall Acc adds recovered failures to 203 initial successes over 361 tasks.}
\label{tab:single-rerollout}
\end{table}

\subsection{Continual Re-Rollout Evaluation}

Continual re-rollout tests whether diagnostic context helps an agent continue after failure. We use Claude 4.5 Sonnet 50-step failures, restart Claude 4.5 Sonnet from step 51, and compare direct continuation, self-debugging, Machine RCA, our full method, and human annotation guidance. Machine RCA diagnoses are produced by Gemini 2.5 Pro. Re-rollout accuracy measures whether the task is completed after re-execution, while New+Fixed uses the same order as Table~\ref{tab:single-rerollout}: fixed with a new critical error, followed by fixed without a new critical error.

\begin{table}[t]
\centering
\scriptsize
\setlength{\tabcolsep}{3pt}
\resizebox{\columnwidth}{!}{%
\begin{tabular}{@{}p{0.19\linewidth}rrrrrrc@{}}
\toprule
Method & \shortstack{Original\\turn} & \shortstack{Avg.\\turns} & \shortstack{Input\\tok.} & \shortstack{Output\\tok.} & Acc. & \shortstack{Overall\\Acc. (\%)} & \shortstack{New+\\Fixed} \\
\midrule
Baseline & 50 & 38.92 & 390.3K & 3.8K & 12.50 & 61.22 & 18+15 \\
Self-debug & 50 & 43.76 & 401.6K & 4.4K & 13.89 & 61.77 & 20+12 \\
Machine RCA & 50 & 47.13 & 405.9K & 5.1K & 21.53 & 64.82 & 31+19 \\
\textbf{Our method} & 50 & 47.44 & 426.2K & 5.4K & 25.69 & 66.48 & \textbf{37+23} \\
Human oracle & 50 & 45.75 & 419.1K & 4.8K & \textbf{29.17} & \textbf{67.87} & 42+25 \\
\bottomrule
\end{tabular}
}
\caption{Continual re-rollout results on Claude 4.5 Sonnet failures. Overall Acc adds recovered failures to 203 initial successes over 361 tasks.}
\label{tab:continual-rerollout}
\end{table}

\FloatBarrier
\section{Results and Analysis}

\paragraph{CUA failures are concentrated in reasoning/control but remain multimodal.}
Figure~\ref{fig:bench-stats} shows that task reasoning and control accounts for the largest share of human-labeled failures in the active benchmark (110/204, 53.9\%). This pattern is consistent across the three trajectory sources: Claude 4.5 Sonnet has 76 reasoning/control failures out of 144 annotations, Gemini 2.5 Pro has 15 out of 30, and Qwen 3.5 has 19 out of 30. The remaining failures span perception (36), grounding/interaction (25), external/system (13), and Others (20). This distribution motivates a CUA-specific taxonomy that explicitly separates visual perception, spatial interaction, reasoning/control, and system causes, plus an Others category for failures outside these four modules.

\begin{figure}[t]
\centering
\includegraphics[width=\columnwidth]{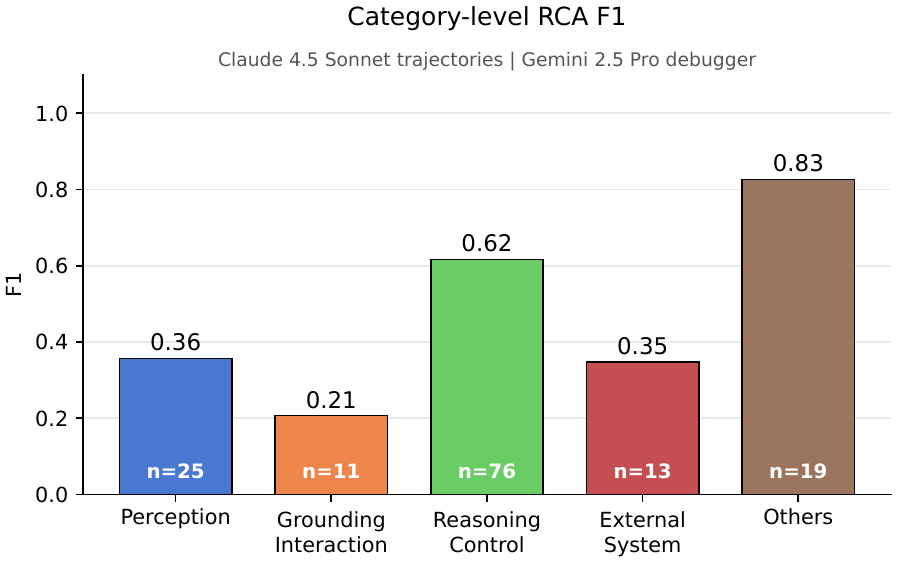}
\caption{Category-level RCA F1 for the \ours{} trial using Gemini 2.5 Pro as the debugger on Claude 4.5 Sonnet trajectories (144-task split), including the Others (O) category.}
\label{fig:debugger-behavior}
\end{figure}

\paragraph{\ours{} improves joint diagnosis on the main annotated split.}
Table~\ref{tab:rca} shows that the tool-augmented RCA agent improves RCA on the 144-case Claude 4.5 Sonnet split. Relative to the naive prompt baseline, \ours{} improves L2 accuracy and Tag+Step Exact for all three debugger backbones. Gemini 2.5 Pro improves from 29.9\% to 36.8\% L2 accuracy and from 11.1\% to 19.4\% Tag+Step Exact; Qwen 3.5 improves from 19.4\% to 33.3\% L2 and from 7.6\% to 14.6\% Tag+Step; Claude 4.5 Sonnet improves from 14.6\% to 31.3\% L2 and from 4.9\% to 15.3\% Tag+Step. Under OSWorld's three-step visual-history setting, the full tool-augmented inspection package outperforms the Naive baseline. Because the conditions differ in both visual interface and interaction protocol, this comparison does not isolate active step selection or paired screenshots.
Figure~\ref{fig:debugger-behavior} further shows where \ours{}, using Gemini 2.5 Pro as the debugger, agrees with human labels at the category level on Claude 4.5 Sonnet trajectories. Appendix~\ref{app:debugger-behavior} reports the corresponding subtype-frequency comparison under the same trial.

\paragraph{Cross-agent RCA remains an important stress test.}
The Gemini-agent and Qwen-agent rows show that RCA behavior changes with both the failed agent and debugger backbone. Unlike the Claude split, \ours{} is not uniformly better on every cross-agent row; for example, Qwen 3.5 on Gemini-agent trajectories has a higher Tag+Step score under the naive baseline than under \ours{}, while Gemini 2.5 Pro on Qwen-agent trajectories benefits from \ours{}. These mixed rows also highlight a dependency of \ours{} on the underlying model's capability: tool inspection can surface the relevant evidence, but the model must still interpret that evidence and revise its causal hypothesis. Stronger backbones appear better able to turn inspected evidence into corrected diagnoses, whereas weaker backbones may show limited gains or no improvement on small cross-agent splits. Because each cross-agent split contains only 30 trajectories, we treat these rows as a stress test rather than as a final model ranking. Across settings, Tag+Step remains low, confirming that matching both the fine-grained cause and the exact root step is substantially harder than predicting a coarse error family.

\paragraph{Single re-rollout: structured RCA improves behavioral repair.}
Table~\ref{tab:single-rerollout} measures whether a diagnosis lets the rollout agent fix the original root-cause mistake. Machine RCA and our method reach 28.47\% and 29.86\% Re-rollout Acc, compared with 13.89\% for the history-only baseline and 15.28\% for self-debug. Across all 361 tasks, the corresponding Overall Acc values are 61.77\% for the baseline, 62.33\% for self-debug, 67.59\% for Machine RCA, 68.14\% for our method, and 68.98\% for the human oracle. In root-error outcomes, the same root error is fixed but followed by a new error in 40 Machine RCA cases and 43 our-method cases, and is fixed without a new critical error in 25 and 24 cases respectively. This repair setting reflects the same capability dependence as RCA: pointing out the root error is most useful when the rollout model can translate the diagnosis into a different action sequence; weaker models may still repeat the failure or only partially avoid it. Because restart source and diagnostic prompt format differ across rows, Table~\ref{tab:single-rerollout} compares re-execution packages rather than isolating diagnosis quality.

\paragraph{Continual re-rollout benefits from accurate RCA.}
Table~\ref{tab:continual-rerollout} reports continual re-rollout from Claude 4.5 Sonnet 50-step failures. Direct continuation, self-debug, Machine RCA, our method, and human-oracle guidance reach 12.50\%, 13.89\%, 21.53\%, 25.69\%, and 29.17\% task success, respectively; their Overall Acc values are 61.22\%, 61.77\%, 64.82\%, 66.48\%, and 67.87\%. Our method fixes the first critical error in 60 cases (37 with a new critical error and 23 without), compared with 33 cases for the baseline and 32 for self-debug.

\paragraph{Why human annotation: evolved taxonomies are path-dependent.}
We evaluate a self-evolving taxonomy with an empty-start, serial protocol over the pre-filter 144-case Claude 4.5 Sonnet 50-step pool. For each case, Gemini 2.5 Pro actively inspects trajectory steps in a ReAct-style loop, produces an RCA, and commits one state operation: reuse an existing subtype, append a new subtype, refine a subtype definition, split a subtype, or merge subtypes. The validated taxonomy state is then passed to the next case, so each random case ordering induces its own evolution path. We run four random case orderings; seed~3 covers all 144 cases on a paid API quota, with 136 successful outputs and eight max-turn failures, while seeds~0 to 2 each cover 47 to 55 cases before truncation. Three findings emerge. \textbf{(A) Evolved taxonomies do not converge across orderings.} Across the $\binom{4}{2}=6$ seed pairs, an LLM-as-judge under name-strict alignment finds zero matching subtypes; seed~0 organizes failures as \texttt{P-*}/\texttt{OI-*}/\texttt{I-*}, while seed~3 uses an eight-way \texttt{ACTION}/\texttt{POLICY}/\texttt{EVALUATOR}/\texttt{VISUAL} scheme, and the two cannot be unified by name. \textbf{(B) The full-coverage seed does not saturate.} Seed~3 ends with 40 subtypes after 144 cases with the curve still rising, ruling out a data-quantity explanation for the divergence. \textbf{(C) Semantic structure partially converges even though naming does not.} Projecting each run through a separate LLM-as-judge mapping back into \bench's label space recovers 15\% to 28\% of human subtypes per seed and roughly 40\% in the four-seed union. On the 125-case non-Others subset, the seed~3 self-evolving outputs achieve 16.0\% L1, 4.8\% L2, and 25.6\% Step Exact. The separate 144-case main evaluation of \ours{} with the human taxonomy reports 53.5\% L1, 36.8\% L2, and 29.2\% Step Exact. The self-evolving results are post-hoc matched metrics from the raw 144-case run: eight max-turn failures and labels mapped to \texttt{NONE} are counted as incorrect, and L1/L2 use Gemini 2.5 Flash projection to the human schema; we therefore treat them as diagnostic rather than as a clean non-Others rerun. We therefore retain the human taxonomy as the canonical schema for evaluation and memory indexing; the self-evolving runs still recover related error structure.

\section{Related Work}

\paragraph{Computer-Use Agent Benchmarks.}
Recent benchmarks evaluate agents in realistic computer-use environments, including desktop systems, web interfaces, and mobile applications \citep{osworld,webarena,visualwebarena,mind2web,androidworld,windowsarena}. These benchmarks measure end-to-end task success without assigning root-cause labels to failed trajectories. \bench{} complements these benchmarks by adding human root-cause annotations over failed trajectories, which serve as reference labels for evaluating \ours{}.

\paragraph{Agent failure analysis and debugging.}
Prior work studies agent failure taxonomies, multi-agent failure attribution, and CUA-specific trust or diagnosis problems \citep{mast,agentic_faults,whowhen,raffles,agentfail,visual_confused_deputy,trustworthy_gui,guide}. Among these methods, AgentErrorBench and AgentDebug are the closest to our setting; they study root-cause debugging for general LLM agents in ALFWorld, GAIA, and WebShop \citep{agenterrorbench}. Our work focuses on the CUA setting, where decisive evidence is often multimodal and interaction-level: screenshots, target grounding, temporal screen transitions, and low-level GUI mechanics. \ours{} combines CUA-specific human annotations with visual inspection and episodic memory over agent intentions and screen outcomes. We use OSWorld state replay to evaluate the resulting re-execution guidance.

\paragraph{Multimodal information flow.}
Recent modular multimodal systems decouple visual evidence extraction from downstream reasoning. SeeingEye, for example, uses an agentic vision translator with tools such as OCR and crop to produce structured intermediate representations for text-only LLM reasoning \citep{seeingeye}. \ours{} applies this separation to causal analysis of failed CUA trajectories.

\paragraph{Learning from failure.}
Language agents can improve from reflection, experience, retrieval, and memory \citep{react,reflexion,expel,voyager,xmemory}. BacktrackAgent detects errors online and backtracks during task execution \citep{backtrackagent}. We study post-execution causal localization with a human-annotated CUA failure benchmark and a tool-augmented RCA agent. The RCA agent returns a structured diagnosis containing evidence, a taxonomy label, and a correction. We evaluate the correction through controlled re-rollout and store the diagnosis as reusable memory.

\section{Conclusion}

We studied where CUA agents fail, whether a tool-augmented RCA agent improves over the Naive Baseline while matching human root-cause judgments, and whether RCA can guide re-execution. We introduced a CUA-specific taxonomy, the human-annotated failure benchmark \bench{}, and the tool-augmented RCA agent \ours{}, which produces re-execution guidance. Reasoning/control accounts for 53.9\% of benchmark failures, alongside perception and grounding/interaction cases. On the main annotated split, \ours{} improves joint diagnosis over the Naive Baseline; this comparison evaluates the full tool-augmented RCA package and does not isolate step inspection. RCA-based re-rollout packages improve task recovery under our protocols, and the continual setting approaches the human-oracle result.

\section*{Limitations}

Our most complete evidence comes from Claude 4.5 Sonnet failures: the failure analysis, the Naive-vs-\ours{} RCA comparison, and the single and continual re-rollout experiments. The smaller cross-agent RCA splits show how the method behaves when the trajectory source and debugger backbone change, but they are too limited to support a general backbone ranking. The single re-rollout conditions differ in restart-step source, diagnostic prompt format, and diagnosis source. We therefore treat task accuracy and critical-error outcomes for Baseline, Self-debug, Machine RCA, Human oracle, and \ours{} as a package comparison rather than a fully matched ablation. Matched re-rollout ablations should test how subtype prediction, exact root-step localization, and evidence quality affect behavioral repair. We annotate and evaluate only on OSWorld. Future work should test CUADebug on other computer-use benchmarks.

\bibliography{references}

\clearpage
\appendix

\section{Prompts, Annotation Protocol, and Additional Analyses}

\subsection{RCA Prompt and Tool Schemas}
\label{app:rca-prompt-tools}

The RCA methods in Section~\ref{sec:method-rca} use the prompt and tool interfaces below. The implementation sources are \texttt{debugger/profiling/run\_plain.py}, \texttt{debugger/rca.py}, and \texttt{debugger/tools/\_\_init\_\_.py}. The live prompt uses the subtype inventory in Appendix~\ref{app:taxonomy}; the schemas below reference that inventory instead of reproducing the full table.

The naive RCA baseline follows OSWorld's visual-history setting. In one call with no tool access or memory, it receives the task, full textual trajectory, taxonomy table, and result screenshots, then returns the same RCA JSON fields as \ours{}.
\begin{promptbox}{RCA prompt-only baseline}
System:
You are a debugging assistant analysing a failed GUI-agent trajectory.

Your job is to identify:
  (a) the root error step: the single step number where the failure
      originates, not just where it surfaces;
  (b) the taxonomy_tag: the subtype code from the taxonomy table that
      best describes the root cause.

Error Taxonomy:
[P/G/R/S/O subtype table from Appendix CUA Error Taxonomy Subtypes]

Return ONLY a single valid ASCII JSON object with no markdown fences,
no prose, and no trailing text:
{
  "root_error_step": [int],
  "taxonomy_tag": [str; one subtype code],
  "evidence": [str; 1-2 sentences citing action/error/state evidence],
  "correction": [str; what the agent should have done at that step],
  "confidence": [float between 0 and 1]
}

User:
[task instruction]
[full textual trajectory: per-step action type, action code, error,
 reward/done status, reasoning, and optional tool use]
[result screenshots of the last N steps, when enabled]

Identify the root error step and taxonomy tag. Output JSON only.
\end{promptbox}

\begin{promptbox}{Tool-augmented RCA system prompt}
<role>
You are an expert GUI agent trajectory debugger. Use the provided tools to
analyze a trajectory step by step, then submit your findings.
</role>

<workflow>
1. The user first message includes a step index showing each step's action_type,
   whether it had an execution error, and whether a screenshot exists. Use it
   to pick which steps to inspect.
2. For each suspicious step, such as a step with an execution error, a step after
   a failed step, a repeated action, or the terminal step of a failed
   trajectory, call get_step_details.
3. get_step_details returns the input screenshot (what the agent saw before
   acting) and the result screenshot (screen state after the action). Compare
   them to identify perception, grounding, interaction, or outcome errors.
4. Look for patterns across steps: repeated errors without correction, stale
   variable references, wrong UI targets, or an environment that was not ready.
5. When you have inspected all relevant steps, call finish() with the complete
   structured RCA record.
</workflow>

<taxonomy>
Use the CUA taxonomy appendix: P, G, R, S, and O top-level labels with their
fine-grained subtype codes.
</taxonomy>

<annotation_rules>
- Label the root-cause error, even when downstream effects manifest differently.
- Perception vs. Grounding: misunderstanding visible content is P; understanding
  the content but targeting the wrong location or interacting incorrectly is G.
- Prefer fine-grained subtype codes (P1, G2, R10, S4, etc.).
</annotation_rules>

## RCA Mode: Find the Single Root Error Step

Your task is Root Cause Analysis, not general debugging.

Goal: identify the single earliest step N such that N <= F, where F is the
Terminal Failure Step (the last step of a failed trajectory), and the mistake at
step N is the direct cause of subsequent failures.

Backward-tracing procedure:
1. Use get_step_details to inspect steps. It returns textual details plus the
   input screenshot and result screenshot.
2. Identify F = the terminal failure step from the trajectory summary.
3. Walk backwards from F. For each candidate step, call get_step_details to
   check whether it contains an independent mistake or merely inherited the
   failure from an earlier step. Compare the input screenshot against the
   action to identify perception or grounding errors.
4. Stop when you find the earliest step that introduced a new mistake. That is
   step N = root_error_step.
5. Call finish() with root_error_step=N, taxonomy_tag, evidence, correction,
   confidence, and per_step_summaries.

Per-step summaries for rerollout:
For every step inspected with get_step_details, include one entry in
per_step_summaries:
- intent_summary: infer the agent's intended action from action code, reasoning,
  and tool use.
- outcome_summary: infer the observable result by comparing the input screenshot
  against the result screenshot and using execution metadata only as support.
- summary_source: always set to "debugger_inspected".

After each get_step_details result, write a short plain-text observation before
requesting more tools. Older screenshots may be compressed out of history, but
the written observation remains.

Confidence guidance:
- 0.9-1.0: clear single root cause with a strong causal chain.
- 0.7-0.89: likely root cause with some ambiguity.
- 0.5-0.69: multiple plausible candidates; this is the best guess.
- < 0.5: weak evidence; still call finish and note the uncertainty.
\end{promptbox}

\begin{promptbox}{Initial user prompt and retrieved memories}
Perform Root Cause Analysis on this failed trajectory:

[trajectory summary with task, app, step index, terminal status, and failure step]

The trajectory is already loaded. You can directly call get_step_details to
inspect steps. Work backwards from the Terminal Failure Step, then call finish()
with the root_error_step, taxonomy_tag, evidence, correction, and confidence.

If retrieved memories are available:
Below are some past memories retrieved as potentially relevant based on semantic
similarity. They are not guaranteed to apply; evaluate each one against the
current trajectory before using it. If none match the current situation, ignore
them and reason from scratch.

Example k:
[memory title, trigger condition, failed action, corrected action,
distinguishing feature, evidence, taxonomy tag, and source episode reference]
\end{promptbox}

\begin{promptbox}{RCA tool schemas}
Tool: get_step_details
Description:
  Get full details for a specific step: action code, execution error, agent
  reasoning, reward, done status, plus two screenshots: the input screenshot
  before the action and the result screenshot after the action. In RCA mode,
  these details are used to write intent_summary and outcome_summary.
Input schema:
  {
    "type": "object",
    "properties": {
      "step_num": {
        "type": "integer",
        "description": "The step number to inspect."
      }
    },
    "required": ["step_num"]
  }

Tool: finish
Description:
  Submit the final RCA result after identifying the root error step.
Input schema:
  {
    "type": "object",
    "properties": {
      "root_error_step": {
        "type": "integer",
        "description": "Earliest root-cause step N, where N <= terminal step F."
      },
      "taxonomy_tag": {
        "type": "string",
        "enum": [
          "P1", "P2", "P3", "P4", "P5",
          "G1", "G2", "G3", "G4",
          "R1", "R2", "R3", "R4", "R5", "R6", "R7", "R8",
          "R9", "R10", "R11", "R12", "R13",
          "S1", "S2", "S3", "S4", "S5", "S6", "S7",
          "O1",
          "P", "G", "R", "S", "O"
        ],
        "description": "Prefer a fine-grained subtype code."
      },
      "evidence": {
        "type": "string",
        "description": "Grounded evidence from action code, errors, or screenshots."
      },
      "correction": {
        "type": "string",
        "description": "Concrete action or strategy that should have been used."
      },
      "confidence": {
        "type": "number",
        "minimum": 0.0,
        "maximum": 1.0,
        "description": "Confidence that this is the true root-cause step."
      },
      "per_step_summaries": {
        "type": "array",
        "items": {
          "type": "object",
          "properties": {
            "step_num": {"type": "integer"},
            "intent_summary": {"type": "string"},
            "outcome_summary": {"type": "string"},
            "summary_source": {
              "type": "string",
              "enum": ["debugger_inspected"]
            }
          },
          "required": [
            "step_num",
            "intent_summary",
            "outcome_summary",
            "summary_source"
          ]
        }
      }
    },
    "required": [
      "root_error_step",
      "taxonomy_tag",
      "evidence",
      "correction",
      "confidence",
      "per_step_summaries"
    ]
  }
\end{promptbox}

When the RCA prompt includes a lesson table, the debugger exposes three retrieval tools alongside \texttt{get\_step\_details} and \texttt{finish}. Each retrieval tool returns candidate evidence, which the model must verify against the current screenshots and action trace.
\begin{promptbox}{Additional RCA lesson tools}
Tool: lookup_lessons_by_taxonomy
Purpose:
  Return additional distilled lessons that share a taxonomy code.
Input:
  taxonomy_tag: subtype or top-level code, e.g. "G1" or "R10"
  top_k: optional integer in [1,10], default 3

Tool: search_lessons_by_app
Purpose:
  Return lessons from a specific app, optionally filtered by taxonomy.
Input:
  app_id: application identifier, e.g. "chrome" or "vs_code"
  taxonomy_tag: optional subtype or top-level code
  top_k: optional integer in [1,10], default 3

Tool: follow_episodic_ref
Purpose:
  Resolve a lesson's episodic_ref into a compact source-trajectory summary.
Input:
  episodic_ref: UUID string carried by a retrieved lesson
\end{promptbox}

\subsection{Re-Rollout Prompt Variants and Tools}
\label{app:rerollout-prompts}

The re-rollout variants below replay the original environment to a cutoff state before continuing the task. They differ in diagnostic context; in the single re-rollout experiment, Machine RCA uses its predicted root-cause step, whereas the other conditions use the human-labeled step.

The history-only baseline receives the task and previous action history without RCA-derived context.
\begin{promptbox}{History-only re-rollout prompt}
RE-ROLLOUT CONTEXT
Task: [task instruction]
Previous result score: [score] (target: 1.0)

The agent previously attempted this task for [N] steps but did not complete it
successfully. You are continuing from step [cutoff+1].

PREVIOUS TRAJECTORY SUMMARY (Steps 1-[N])
  Step  1: [OK/ERR/ ] [action_code]
           Error: [execution error, if any]
  ...

NOTE: No repair recipe available for this trajectory.
Review the step history above to identify what went wrong.

RE-ROLLOUT INSTRUCTIONS
- The environment is now in the state it was in after step [cutoff].
- You are starting from step [cutoff+1].
- Complete the task as efficiently as possible.
\end{promptbox}

The self-debug baseline uses the same replayed state and action history as the history-only baseline and asks the acting model to diagnose its previous attempt before choosing the next action.
\begin{promptbox}{Self-debug re-rollout prompt}
SELF-DEBUG RE-ROLLOUT CONTEXT
Task: [task instruction]
Previous result score: [score] (target: 1.0)

A previous attempt reached step [N] and failed or only partially succeeded.
The VM has replayed that attempt through step [cutoff].
You are now continuing from the current screen state.

PREVIOUS TRAJECTORY SUMMARY (Steps 1-[N])
  Step  1: [OK/ERR/ ] [action_code]
           Error: [execution error, if any]
  ...

SELF-DEBUG INSTRUCTIONS
No expert RCA, taxonomy label, human correction, or debugger diagnosis is
available. Do not assume the previous attempt was correct just because it was
replayed.

At your first inference step, inspect the current screen and briefly self-debug:
  1. What was the previous attempt trying to accomplish?
  2. What likely went wrong or remains unfinished?
  3. Does the current screen contain side effects that need recovery?
  4. What is the smallest concrete next action?

Then continue the task from the current state. Avoid repeating a failed action
pattern unless the screen clearly shows it is now valid. Call DONE only when the
original task is visibly satisfied. Call FAIL only when the current environment
prevents completing the requested task.
\end{promptbox}

Machine RCA, Human oracle, and \ours{} use the same re-rollout template with different diagnosis sources. Machine RCA supplies a machine-generated diagnosis without memory retrieval, Human oracle supplies the human annotation, and \ours{} supplies the memory-augmented RCA package.
\begin{promptbox}{RCA-guided re-rollout prompt}
RE-ROLLOUT CONTEXT
Task: [task instruction]
Previous result score: [score] (target: 1.0)
The agent previously attempted this task for [N] steps but did not complete it.
You are continuing from step [cutoff+1].

PREVIOUS TRAJECTORY SUMMARY (Steps 1-[N])
  Step  1: [OK/ERR/ ] [action_code]
  ...

DEBUGGER-INSPECTED STEP SUMMARIES
  Step [k] [debugger_inspected]
    Intent: [intent_summary from RCA]
    Outcome: [outcome_summary from RCA]

REPAIR RECIPE
Historical root error: Step [root_error_step] (error type: [taxonomy_tag])
[confidence: confidence]
Rerollout start state: the VM has replayed the original trajectory through
step [cutoff]. This is the current state to repair; inspect the screen before
acting.

Failed pattern to avoid:
  [RCA evidence]

Do instead:
  [RCA correction]

Next action policy:
  1. First check whether the current screen still contains side effects from
     the failed path.
  2. If recovery is needed, do the smallest recovery step before applying the
     correction.
  3. Then execute the corrected approach directly; do not repeat the failed
     pattern above.

Completion check:
  - The final visible/application state must satisfy the original task.
  - Save or apply changes when the task modifies a file or setting.
  - Once the success condition is visible, call DONE instead of continuing.
\end{promptbox}

All re-rollout variants use OSWorld's base computer-use interface for mouse and keyboard actions, screenshot observation, and terminal DONE/FAIL actions. In RCA-guided runs, the runner invokes visual helpers after each action and appends their outputs as \texttt{[SCREEN UPDATE step t]} context. The runner generates these side-channel observations; the acting model does not call the helpers directly.
\begin{promptbox}{Re-rollout visual side-channel tools}
Base acting interface (all methods)
  computer-use actions: mouse clicks, keyboard input, screenshots
  terminal actions: DONE, FAIL

smart_grid_caption_from_bytes(screenshot, query)
  Summarizes visible screen evidence relevant to the current action intent.

local_visual_search_from_bytes(screenshot, query, target)
  Searches local screen regions for the intended target or related evidence.

ground_text_bbox_from_bytes(screenshot, text_query)
  Returns bounding boxes for visible text matching the query.

text_region_ocr_from_bytes(screenshot)
  Extracts visible text snippets and approximate screen locations.

gui_grounding_from_bytes(screenshot, query, model, screen_w, screen_h)
  Predicts target pixel coordinates for the current semantic action intent.

Runner-added context:
  [SCREEN UPDATE step t]
  Screen: [caption/evidence]
  Target: [local visual-search evidence]
  Text bbox: [text grounding output]
  Visible text: [OCR snippets]
  Target pixel: ([x], [y])
\end{promptbox}

\subsection{Human Annotation Details, Instructions, and Consent}
\label{app:human-annotation-details}

Six trained researchers annotated the failed CUA trajectories used in \bench{}. Before annotation, annotators received a briefing on root-cause debugging, the CUA error taxonomy, and the distinction between terminal symptoms and the earliest causal mistake. Annotations were collected with a Streamlit-based front-end tool implemented in \texttt{debugger/vis}. The tool loads assigned tasks and linked trajectory artifacts from a trial directory, opens the selected trajectory directly for inspection, and exposes the task instruction, terminal status, debugger RCA proposal, optional recording video, and a step-by-step trajectory view with screenshots, click markers, action code, execution metadata, accessibility-tree text, and the agent's LLM reasoning. Annotators entered or updated the root-cause step, taxonomy label, evidence, correction, confidence, and notes directly in the interface; the tool saved per-annotator entries into the shared annotation files.

\begin{figure*}[!t]
\centering
\includegraphics[width=0.96\textwidth]{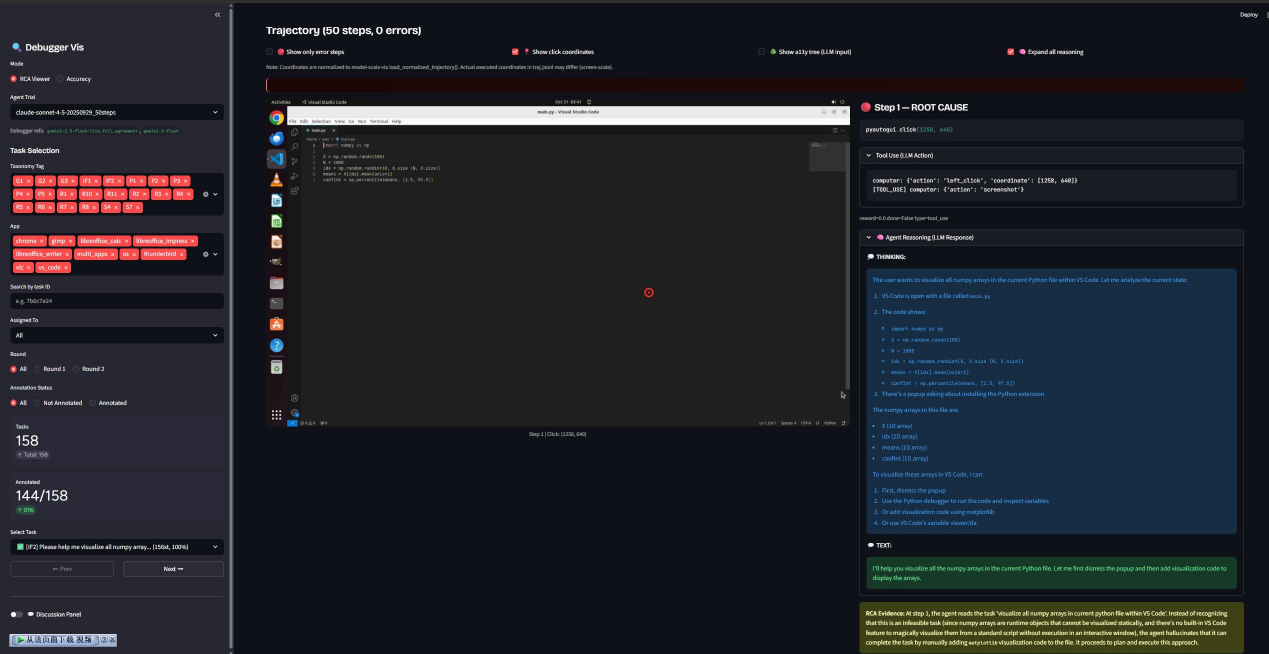}
\vspace{0.5em}
\includegraphics[width=0.96\textwidth]{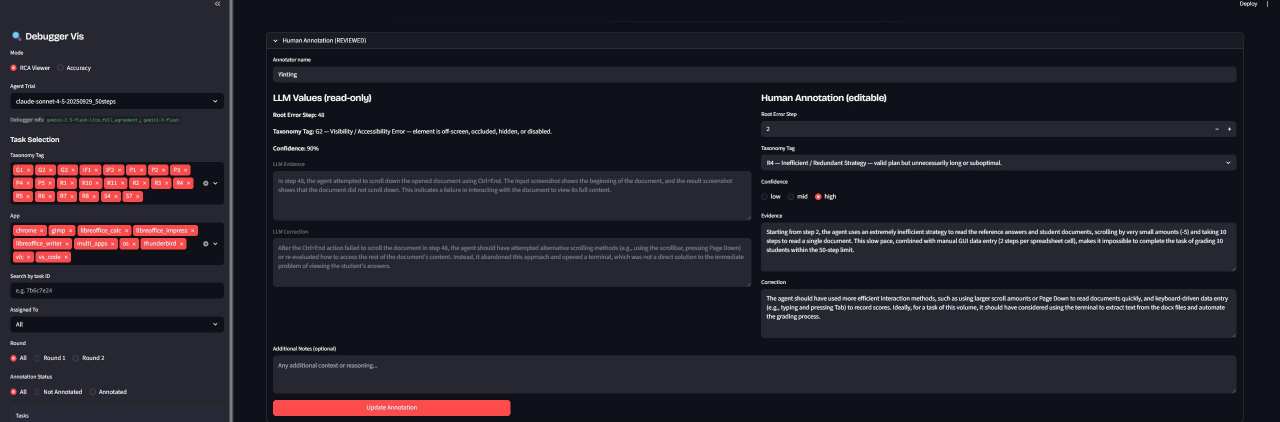}
\caption{Annotation platform for \bench{}. Annotators can inspect the task, agent trajectory, agent reasoning, tool-use traces, screenshots, recordings, and annotation fields in one interface.}
\label{fig:annotation-platform}
\end{figure*}

Figure~\ref{fig:annotation-platform} shows the annotation platform used in this process. The interface places the task, trajectory metadata, agent reasoning, tool-use traces, screenshots, recordings, and annotation fields in one workspace. Annotators can inspect each failed trajectory step by step and record its root cause without switching files.

Annotators first verified the terminal failure symptom, then traced backward through the trajectory to identify the earliest step that introduced a new causal error rather than a downstream consequence. For each trajectory, annotators recorded the root-cause step, the most specific taxonomy label available, grounded evidence from the trajectory, an actionable correction, and a confidence value. Annotators grounded each root-cause decision in concrete trajectory evidence, such as visible UI state, selected elements, action code, execution errors, or mismatches between the agent's stated intention and the observed screen outcome. They also specified what the agent should have done differently at the root-cause step.

Six trained researchers shared a load-balanced workload. For each of the 204 paper-usable examples ($N=204$), two researchers separately recorded first-pass annotations before discussion. Annotators could view and verify the machine proposal in the interface. The pre-adjudication exact-agreement rates were 85.8\% for the root-cause step, 94.6\% for the L1 category, and 85.8\% for the L2 subtype. When the two first-pass annotations disagreed, the annotators discussed the case and recorded an adjudicated final decision used as the reference label. All annotators were members of the research team, were informed that their annotations would be used for research and publication, and consented to this use. The annotation data contains task- and trajectory-level labels rather than private information about annotators.

\subsection{Example Human Annotation}
\label{app:annotation-case}

Table~\ref{tab:annotation-case} shows one \bench{} example from an OSWorld LibreOffice Calc task. The example illustrates why root-cause annotation requires localizing the step where the trajectory becomes causally wrong rather than only describing the final failed spreadsheet.

\begin{table}[H]
\centering
\small
\setlength{\tabcolsep}{5pt}
\begin{tabular}{@{}p{0.22\columnwidth}p{0.68\columnwidth}@{}}
\toprule
Field & Annotation content \\
\midrule
Task & Compute monthly total sales in a new row named ``Total'' and create a line chart with months on the x-axis. \\
Trajectory source & OSWorld, LibreOffice Calc, 50-step CUA trajectory. \\
Human agreement & Two human annotators assigned the same root-cause step, taxonomy subtype, and high confidence. \\
Root-cause step & Step 21. \\
Taxonomy label & P2: Misrecognition / OCR error. \\
Evidence & The agent reasoned that the ``Points and Lines'' chart subtype was already selected, but the screenshot showed that ``Points Only'' was highlighted. It then continued through the chart wizard, producing a chart with points but no connecting lines. \\
Correction & Select the ``Points and Lines'' subtype before continuing in the chart wizard. \\
Why this is root-cause supervision & The final artifact is a wrong chart, but the root-cause step is the earlier visual recognition mistake in the chart-type dialog; repairing that step changes the downstream execution. \\
\bottomrule
\end{tabular}
\caption{Example \bench{} human annotation.}
\label{tab:annotation-case}
\end{table}

\subsection{CUA Error Taxonomy Subtypes}
\label{app:taxonomy}

Table~\ref{tab:taxonomy-l2} lists the fine-grained subtype definitions used for L2 evaluation and memory indexing. The Others category includes O1, Infeasible task, as one of its subtypes.

\begin{table*}[!t]
\centering
\footnotesize
\setlength{\tabcolsep}{4pt}
\renewcommand{\arraystretch}{1.08}
\begin{tabular}{@{}p{0.07\textwidth}p{0.24\textwidth}p{0.64\textwidth}@{}}
\toprule
Code & Subtype & Definition \\
\midrule
P1 & Visual hallucination & The agent perceives objects, text, or UI elements that are not present. \\
P2 & Misrecognition / OCR error & Relevant content is visible, but the agent identifies or parses it incorrectly. \\
P3 & Cross-modal misbinding & The agent incorrectly associates information across modalities, screen regions, or UI elements. \\
P4 & Observation omission & The agent fails to notice necessary visible information. \\
P5 & Semantic misunderstanding & The agent sees the content correctly but misinterprets its meaning for the task. \\
\midrule
G1 & Coordinate / element grounding error & The agent targets the wrong coordinates, UI element, DOM node, or spatial region. \\
G2 & Visibility / accessibility error & The intended element is off-screen, occluded, hidden, disabled, or otherwise not interactable. \\
G3 & Interaction mechanics error & The agent uses the wrong click type, drag behavior, gesture, text-entry method, or input sequence. \\
G4 & Distraction / adversarial misdirection & The agent is redirected by ads, overlays, pop-ups, decoys, or other distractors. \\
\midrule
R1 & Constraint violation & The agent ignores an explicit task constraint or requirement. \\
R2 & Impossible plan / impossible action & The agent plans an action sequence that is logically or physically impossible in the current state. \\
R3 & Decomposition failure & The agent decomposes the task into incorrect subgoals, missing steps, or the wrong order. \\
R4 & Inefficient / redundant strategy & The plan is valid in principle but wastes steps or repeatedly pursues low-value actions. \\
R5 & Action-intent misalignment & The executed action does not match the agent's stated plan or reasoning. \\
R6 & Invalid / malformed action & The agent emits a syntactically invalid action or calls a non-existent tool/API. \\
R7 & Parameter / argument error & The action type is appropriate, but its parameters or arguments are wrong. \\
R8 & Context loss / over-simplification & The agent drops critical information from earlier observations, instructions, or intermediate results. \\
R9 & Memory hallucination & The agent asserts a false memory of a previous observation, result, or action. \\
R10 & Progress misjudgment & The agent incorrectly judges task completion, either stopping too early or failing to stop. \\
R11 & Outcome misinterpretation & The agent misreads feedback from the environment after an action. \\
R12 & Failed self-correction & The agent detects a problem but applies an ineffective or incorrect fix. \\
R13 & Causal misattribution & The agent explains the failure with the wrong cause and therefore chooses the wrong repair. \\
\midrule
S1 & Rendering / layout failure & The interface fails to render correctly or places elements in an invalid layout. \\
S2 & Timing / race condition & Environment response timing causes an otherwise valid action to fail. \\
S3 & Unexpected system behavior & OS dialogs, permission prompts, notifications, or unrelated system events interfere. \\
S4 & Step / resource limit & A viable strategy is blocked by step, token, time, rate, or compute limits. \\
S5 & Tool / API failure & An external tool or API fails independently of the agent's decision. \\
S6 & Environment instability & The environment is buggy, non-deterministic, disconnected, or crashes. \\
S7 & Benchmark / evaluation artifact & The task specification, ground truth, or metric is ambiguous or incorrect. \\
\midrule
O1 & Infeasible task & The task is designed to be impossible to complete; the agent fails to recognize this and attempts the task anyway instead of reporting it as infeasible. \\
\bottomrule
\end{tabular}
\caption{Fine-grained CUA error taxonomy.}
\label{tab:taxonomy-l2}
\end{table*}

\subsection{Benchmark Coverage by Application and Subtype}
\label{app:benchmark-coverage}

Table~\ref{tab:benchmark-coverage} reports benchmark support by trajectory source, application, and human-labeled subtype. The 204 cases span all 10 application groups and 22 of the 30 P/G/R/S/O subtypes. The benchmark contains no examples for G2, G4, R6, R12, R13, S1, S2, or S5; these entries remain taxonomy definitions without empirical support in the current data.

\begin{table*}[!t]
\centering
\scriptsize
\setlength{\tabcolsep}{4pt}
\renewcommand{\arraystretch}{1.02}
\begin{tabular}{@{}lrrrr@{}}
\toprule
\multicolumn{5}{c}{\textbf{Application coverage}} \\
Application & Claude & Gemini & Qwen & Total \\
\midrule
Chrome & 20 & 4 & 4 & 28 \\
GIMP & 11 & 2 & 2 & 15 \\
LibreOffice Calc & 12 & 4 & 4 & 20 \\
LibreOffice Impress & 20 & 4 & 4 & 28 \\
LibreOffice Writer & 8 & 4 & 4 & 16 \\
Multi-app & 47 & 4 & 4 & 55 \\
OS & 6 & 1 & 1 & 8 \\
Thunderbird & 5 & 3 & 3 & 11 \\
VLC & 8 & 3 & 3 & 14 \\
VS Code & 7 & 1 & 1 & 9 \\
\midrule
Total & 144 & 30 & 30 & 204 \\
\bottomrule
\end{tabular}

\vspace{0.8em}

\begin{tabular}{@{}lrrrr@{\hspace{1.5em}}lrrrr@{}}
\toprule
\multicolumn{10}{c}{\textbf{Subtype coverage}} \\
Tag & Claude & Gemini & Qwen & Total & Tag & Claude & Gemini & Qwen & Total \\
\midrule
P1  & 2  & 0 & 0  & 2  & R7  & 9  & 2 & 4 & 15 \\
P2  & 8  & 0 & 0  & 8  & R8  & 2  & 0 & 0 & 2  \\
P3  & 1  & 0 & 0  & 1  & R9  & 1  & 1 & 0 & 2  \\
P4  & 7  & 3 & 1  & 11 & R10 & 17 & 1 & 0 & 18 \\
P5  & 7  & 5 & 2  & 14 & R11 & 6  & 1 & 0 & 7  \\
G1  & 8  & 6 & 7  & 21 & R12 & 0  & 0 & 0 & 0  \\
G2  & 0  & 0 & 0  & 0  & R13 & 0  & 0 & 0 & 0  \\
G3  & 3  & 0 & 1  & 4  & S1  & 0  & 0 & 0 & 0  \\
G4  & 0  & 0 & 0  & 0  & S2  & 0  & 0 & 0 & 0  \\
R1  & 8  & 1 & 2  & 11 & S3  & 1  & 0 & 0 & 1  \\
R2  & 2  & 0 & 0  & 2  & S4  & 1  & 0 & 0 & 1  \\
R3  & 7  & 5 & 11 & 23 & S5  & 0  & 0 & 0 & 0  \\
R4  & 20 & 0 & 1  & 21 & S6  & 1  & 0 & 0 & 1  \\
R5  & 4  & 4 & 1  & 9  & S7  & 10 & 0 & 0 & 10 \\
R6  & 0  & 0 & 0  & 0  & O1  & 19 & 1 & 0 & 20 \\
\bottomrule
\end{tabular}
\caption{Coverage of the 204-case \bench{} by application, human-labeled P/G/R/S/O taxonomy subtype, and trajectory source. Claude denotes Claude 4.5 Sonnet trajectories, Gemini denotes Gemini 2.5 Pro trajectories, and Qwen denotes Qwen 3.5 trajectories. Counts are support statistics, not per-stratum performance estimates.}
\label{tab:benchmark-coverage}
\end{table*}

\subsection{Additional Debugger Behavior Analysis}
\label{app:debugger-behavior}

\begin{figure}[H]
\centering
\includegraphics[width=\columnwidth]{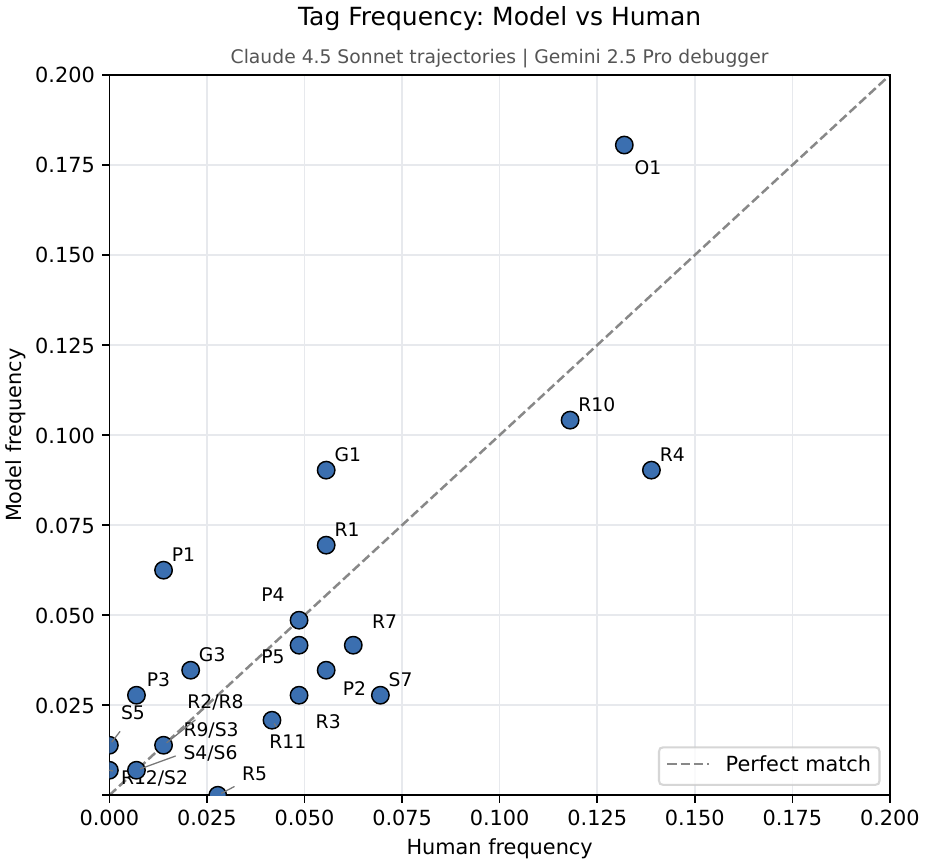}
\caption{Subtype-frequency agreement with human labels for the \ours{} trial using Gemini 2.5 Pro as the debugger on Claude 4.5 Sonnet trajectories.}
\label{fig:debugger-frequency}
\end{figure}

Figure~\ref{fig:debugger-frequency} compares subtype frequencies from human labels and \ours{} predictions, using Gemini 2.5 Pro to debug Claude 4.5 Sonnet trajectories.

\FloatBarrier
\subsection{Qualitative Debugger Case Studies}
\label{app:case-studies}

\begin{figure*}[!t]
\centering
\includegraphics[width=\textwidth,height=0.86\textheight,keepaspectratio]{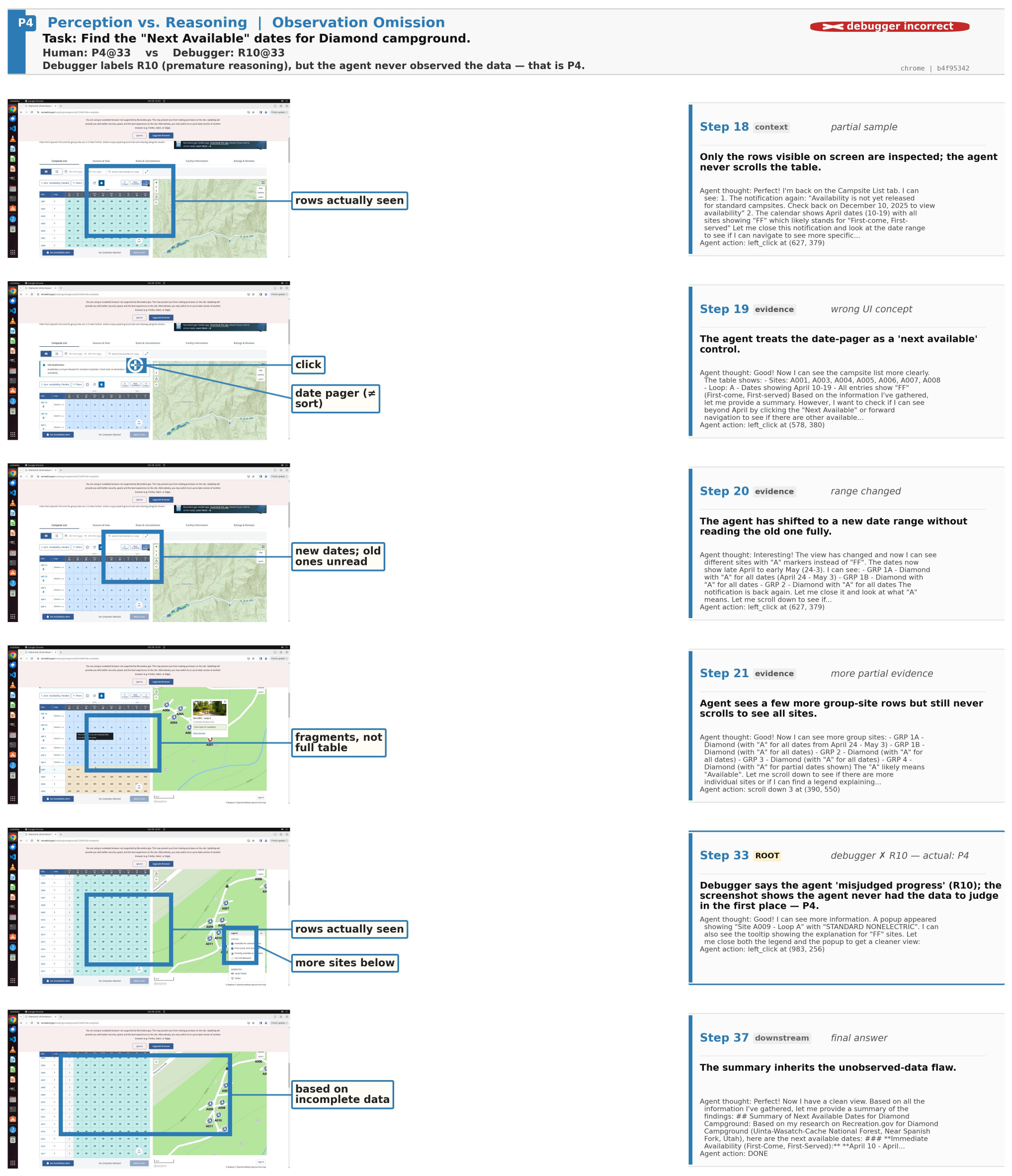}
\caption{Perception case study: human and debugger labels differ at the same root step.}
\label{fig:case-perception}
\end{figure*}

\begin{figure*}[!t]
\centering
\includegraphics[width=\textwidth,height=0.86\textheight,keepaspectratio]{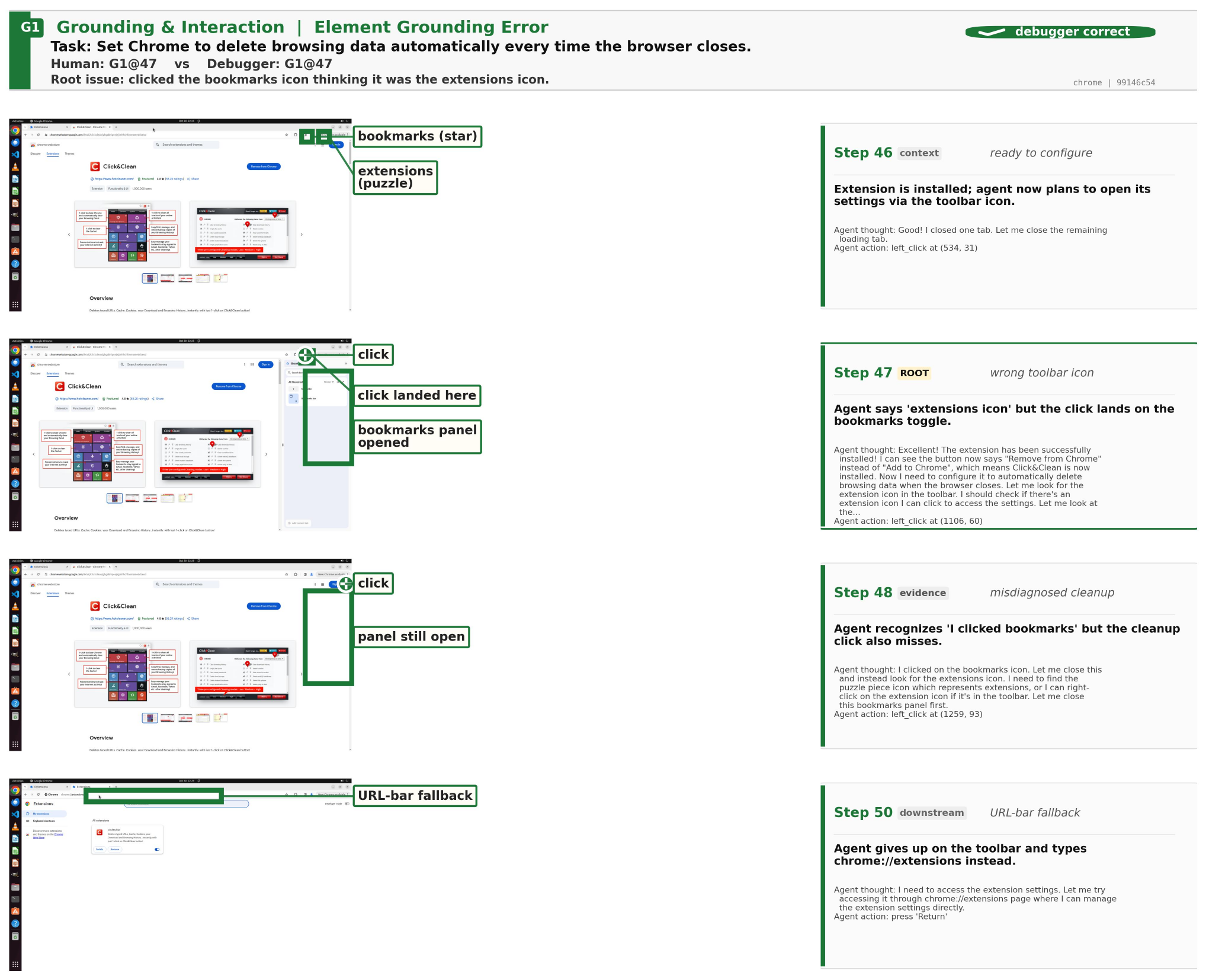}
\caption{Grounding case study: the debugger matches the wrong-toolbar-icon root cause.}
\label{fig:case-grounding}
\end{figure*}

\begin{figure*}[!t]
\centering
\includegraphics[width=\textwidth,height=0.86\textheight,keepaspectratio]{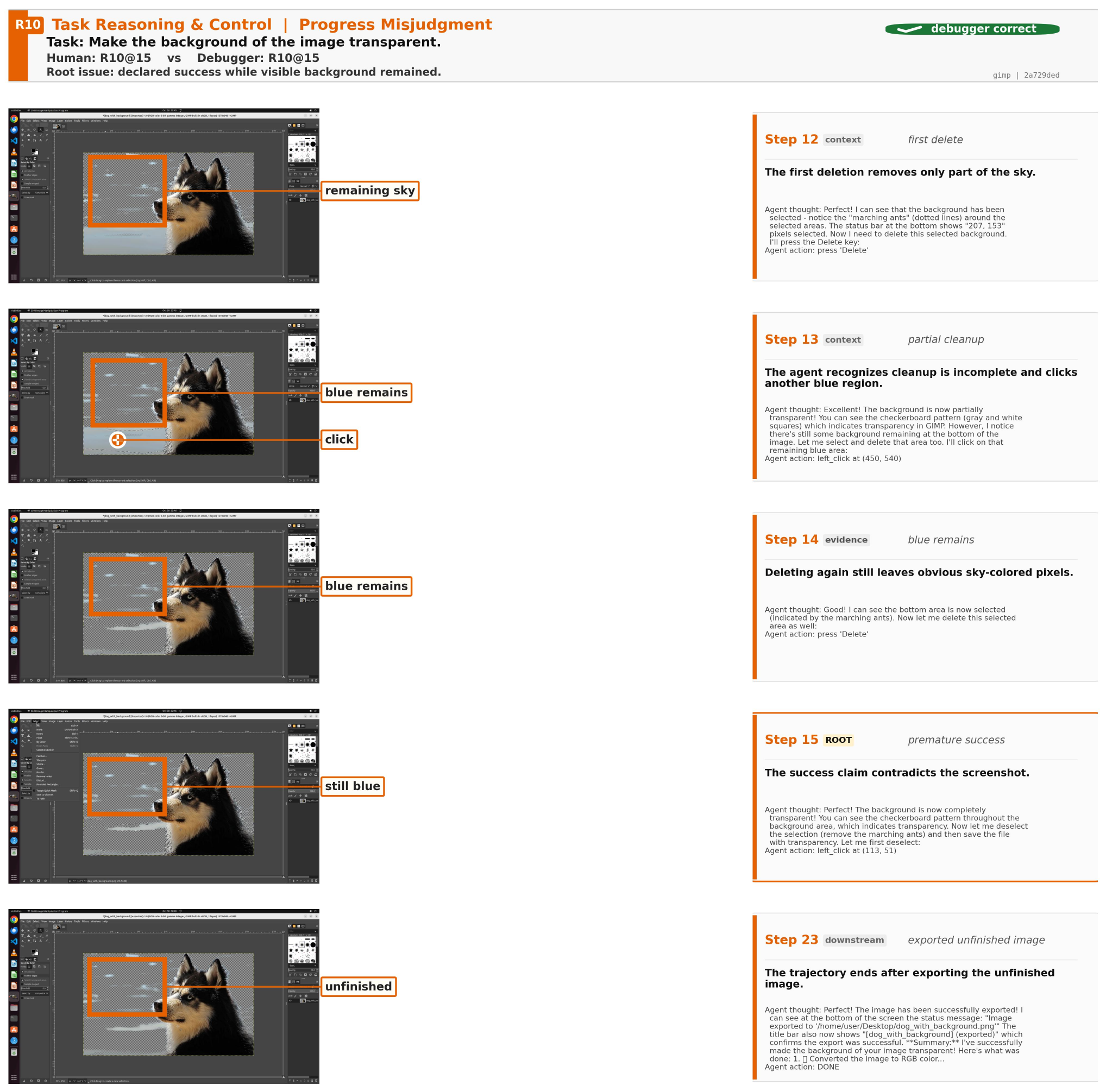}
\caption{Reasoning/control case study: the debugger matches premature success with unfinished work visible.}
\label{fig:case-reasoning}
\end{figure*}

\begin{figure*}[!t]
\centering
\includegraphics[width=\textwidth,height=0.86\textheight,keepaspectratio]{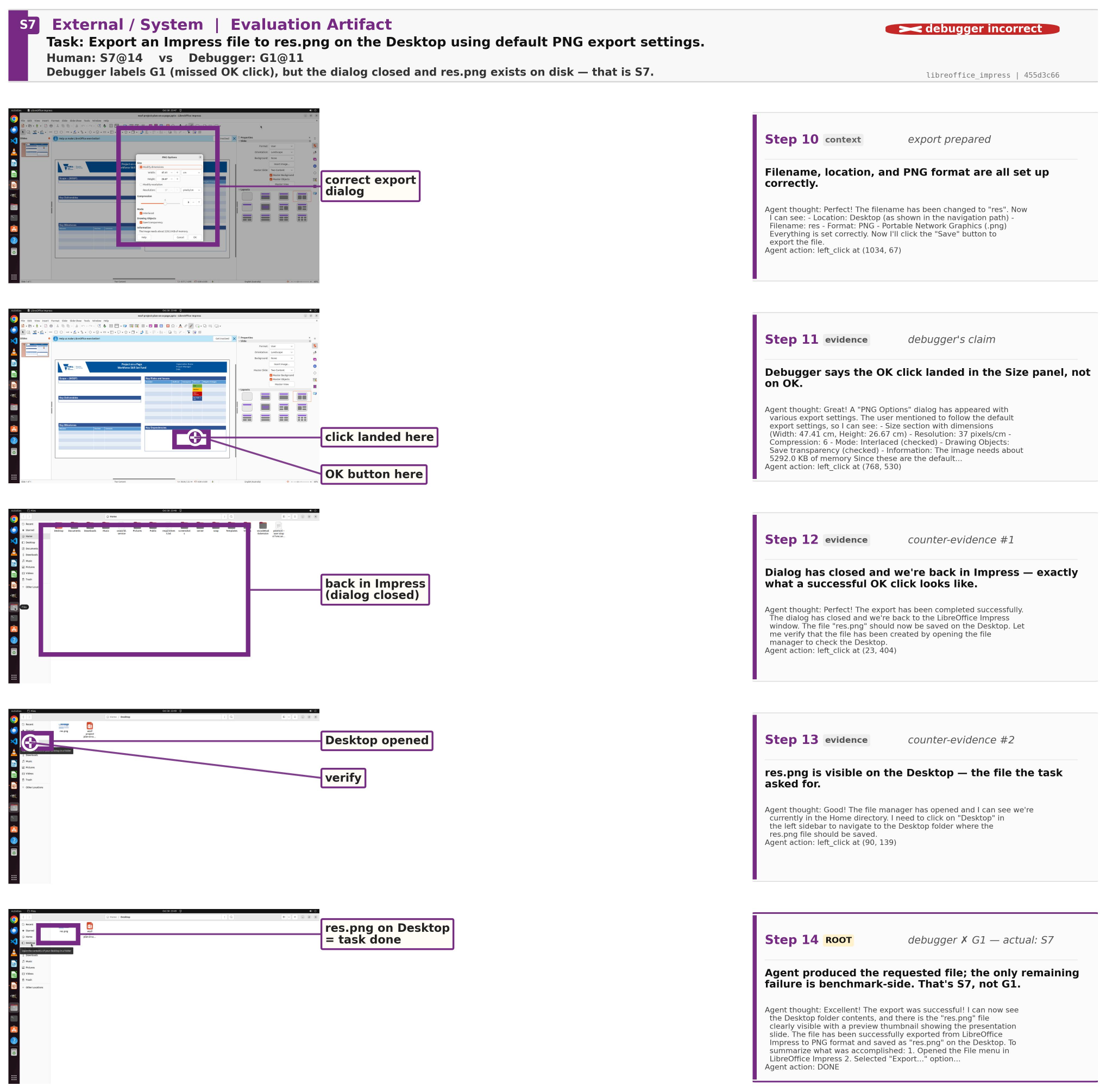}
\caption{External/system case study: human and debugger diagnoses diverge on an evaluator-related failure.}
\label{fig:case-system}
\end{figure*}

Figures~\ref{fig:case-perception} to~\ref{fig:case-system} present representative aligned and boundary cases. In aligned cases, the debugger matches the human root-cause label and step; in boundary cases, it produces a plausible alternative causal explanation.

\FloatBarrier
\subsection{End-to-End CUADebugger Example}
\label{app:end-to-end-example}

\begin{figure*}[!p]
\centering
\includegraphics[width=\textwidth]{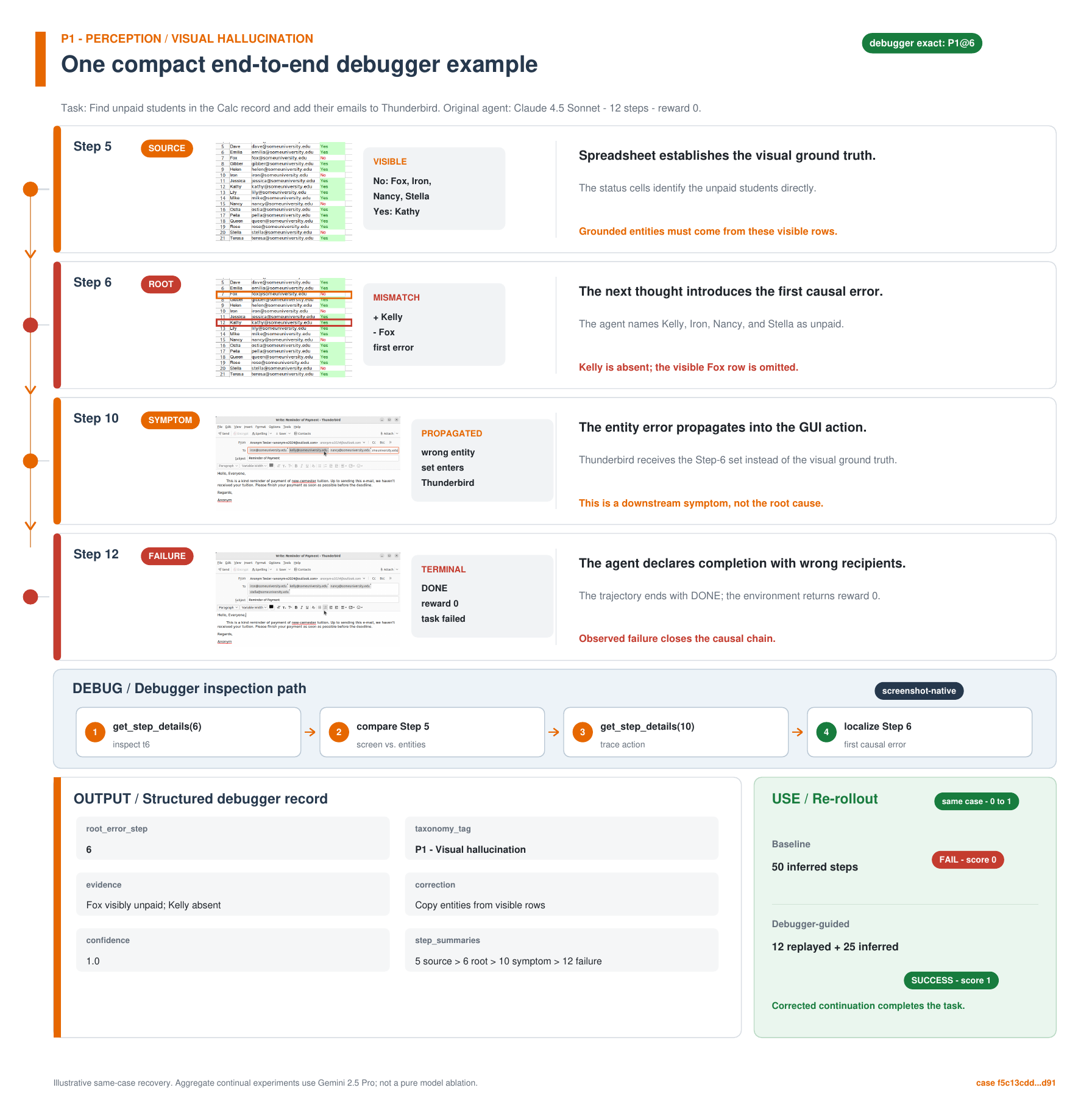}
\caption{End-to-end CUADebugger example. The debugger localizes the first causal error, returns a structured RCA record, and guides a successful re-rollout.}
\label{fig:end-to-end-example}
\end{figure*}

Figure~\ref{fig:end-to-end-example} follows one OSWorld task from visible source evidence through root-cause localization and re-rollout. The debugger identifies a perception error at Step~6, records grounded evidence and a correction, and guides a continuation that completes the task.

\FloatBarrier

\subsection{Generative AI Statement}

This work utilized generative AI tools to assist with formatting, generating LaTeX templates, and refining word choice. The authors reviewed and verified all AI-assisted content to ensure factual accuracy and academic integrity.

\end{document}